\documentclass[aps,reprint,superscriptaddress]{revtex4-1}
\usepackage{amsmath}
\usepackage{bm} 
\usepackage{amssymb}
\usepackage{color}
\usepackage{multirow}
\usepackage[colorlinks,bookmarks=false,citecolor=red,linkcolor=blue,urlcolor=blue]{hyperref}
\usepackage{graphicx}
\usepackage{mathrsfs,bbm}
\usepackage[caption=false]{subfig} 
\usepackage{wrapfig}
\usepackage{physics}
\usepackage{enumerate}
\usepackage[version=3]{mhchem}
\usepackage{gensymb}
\usepackage{dcolumn}
\usepackage{calligra}
\DeclareMathAlphabet{\mathcalligra}{T1}{calligra}{m}{n}
\DeclareFontShape{T1}{calligra}{m}{n}{<->s*[2.2]callig15}{}

\def\sbar{\bar{s}}
\def\tbar{\bar{t}}
\def\Hhat{\hat{H}}

\def\ahat{\hat{a}}

\def\hhat{\hat{h}}

\def\qhat{\hat{q}}
\def\shat{\hat{s}}
\def\that{\hat{t}}
\def\xhat{\hat{x}}
\def\yhat{\hat{y}}

\newcommand{\RomanNumeralCaps}[1]
    {\MakeUppercase{\romannumeral #1}}

\begin{document}
\title{Quantum paramagnetism and magnetization plateaus in a kagome-honeycomb Heisenberg antiferromagnet}
\author{Meghadeepa Adhikary}
\affiliation{School of Physical Sciences, Jawaharlal Nehru University, New Delhi 110067, India.}
\author{Arnaud Ralko}
\affiliation{Institut N\'eel, Universit\'e Grenoble Alpes $\&$ CNRS, 38042 Grenoble, France.}
\author{Brijesh Kumar}
\affiliation{School of Physical Sciences, Jawaharlal Nehru University, New Delhi 110067, India.}
\date{\today} 

\begin{abstract}
A spin-1/2 Heisenberg model on honeycomb lattice is investigated by doing triplon analysis and quantum Monte Carlo calculations. This model, inspired by Cu$_2$(pymca)$_3$(ClO$_4$), has three different antiferromagnetic exchange interactions ($J_A$, $J_B$, $J_C$) on three different sets of nearest-neighbour bonds which form a kagome superlattice. While the model is bipartite and unfrustrated, its quantum phase diagram is found to be dominated by a quantum paramagnetic phase that is best described as a spin-gapped hexagonal-singlet state. The N\'eel antiferromagnetic order survives only in a small region around $J_A=J_B=J_C$. The magnetization produced by external magnetic field is found to exhibit  plateaus at 1/3 and 2/3 of the saturation value, or at 1/3 alone, or no plateaus.  Notably, the plateaus exist only inside a bounded region within the hexagonal-singlet phase. This study provides a clear understanding of the spin-gapped behaviour and magnetization plateaus observed in Cu$_2$(pymca)$_3$(ClO$_4$), and also predicts the possible disappearance of 2/3 plateau under pressure.
\end{abstract} 

\maketitle
\section{Introduction \label{sec:intro}}
Models of interacting quantum spins are essential to our understanding of magnetism in real materials. They come in different forms, and display a variety of phenomena arising from an interplay of competing interactions, quantum fluctuations and lattice geometry~\cite{Mattis,Mila}. Antiferromagnetic spin-1/2 Heisenberg model is a problem of fundamental importance to quantum magnetism, and its physics depends sensitively on the underlying lattice. For instance, on honeycomb lattice with uniform nearest-neighbour interactions, it is known to realise N\'eel order in the ground state~\cite{Reger1989,Fouet2001}. But the same spin-1/2 model on kagome lattice harbours a complex spin liquid ground state~\cite{Yan2011,Iqbal2013,Ralko2018,Lauchli2019}. There are materials that realise spin-1/2 honeycomb~\cite{Kataev2005,Moller2008,Tsirlin2010,Kono2020} or kagome~\cite{Helton2007,Okamoto2009} antiferromagnets. The absence of magnetic order on kagome lattice is due to its frustrated geometry. Such a loss of magnetic order can also be caused on honeycomb lattice by allowing the exchange interactions to compete. It can be so done either by having further neighbour interactions \cite{Lamas2013, Merino2018}, or at the very least, by making the nearest-neighbour interactions non-uniform. In this paper, we take the latter route and consider spins on such a non-uniform honeycomb lattice whose nearest-neigbhour bonds form kagome superlattice. We term it as the `kagome-honeycomb' lattice. 
 
The motivation for the present study comes from the recent experimental studies of \ce{Cu2(pymca)3(ClO4)}~\cite{Okutani2019,Sugawara2017}. This compound is reported to have no magnetic order down to 0.6 K, and to exhibit magnetization plateaus at 1/3 and 2/3 of the saturation value. The basic model applicable to this material is the spin-1/2 Heisenberg model on honeycomb lattice with three different nearest-neighbour antiferromagnetic interactions $J^{ }_A$, $J^{ }_B$ and $J^{ }_C$, as shown in Fig.~\ref{fig:ABChoney}. Note that these three exchange interactions form a kagome superlattice on the underlying honeycomb. The material realizes this kagome superstructure via lattice distortions~\cite{Sugawara2017} (consistent with a theorem on the possible distortions of the honeycomb lattice~\cite{Frank2011}). Thus, we have a  kagome-honeycomb Heisenberg antiferromagnet in \ce{Cu2(pymca)3(ClO4)}. It can also be viewed as a system of hexagons formed by two types of bonds (say, $J^{ }_B$ and $J^{ }_C$) and coupled via the third (say, $J^{ }_A$). This is exactly like some spin-1 kagome systems with antiferromagnetic $J^{ }_B$ and $J^{ }_C$, but ferromagnetic $J^{ }_A$~\cite{Wada1997,Hida2000,Uekusa2000,PG2018}. An early example of a frustrated spin-1/2 Heisenberg model with exact dimer singlet ground state on kagome-honeycomb lattice occurs in Ref.~\cite{BKThesis}. 

In this paper, we study the quantum phase diagram of the spin-1/2 Heisenberg antiferromagnet on kagome-honeycomb lattice by doing triplon analysis and unbiased quantum Monte Carlo (QMC) simulations. The theory of triplon fluctuations and the observables computed by QMC produce mutually consistent results not only qualitatively but also quantitatively. Remarkably, in spite of being bipartite and unfrustrated, this model is found to realise in a large part of the phase diagram a quantum paramagnetic phase, while only a small region around $J_A=J_B=J_C$ corresponds to the N\'eel antiferromagnetic phase. This quantum paramagnetic phase is described well as a spin-gapped hexagonal singlet state. We also investigate this model in an external magnetic field, and find the magnetization plateaus at 1/3 and 2/3 of the saturation value, or only one plateau at 1/3, or no plateau at all. In the phase diagram, the region of existence of the 2/3 plateau is found to occur inside that of the 1/3 plateau, which itself exists inside a bounded region within the hexagonal singlet phase. It clearly affirms that \ce{Cu2(pymca)3(ClO4)} realizes hexagonal singlet ground state. Our estimate of the exchange interactions puts this material inside the two-plateau region but close to the boundary. This leads to an interesting testable prediction that the 2/3 plateau in \ce{Cu2(pymca)3(ClO4)} can be made to disappear, say, by applying pressure. 

This paper is organized as follows: in Sec.~\ref{sec:model} we describe the model and discuss its key qualitative aspects; in Sec.~\ref{sec:trip} we do triplon analysis of the model, and present the quantum phase diagram obtained from it; in Sec.~\ref{sec:num}, we present the results obtained from QMC simulations. Section~\ref{sec:Mag} is devoted to the study of magnetization plateaus, 
with implications for \ce{Cu2(pymca)3(ClO4)}. 
We conclude with a summary and outlook in Sec.~\ref{sec:sum}. 

\section{Model \label{sec:model}} 
\begin{figure}[t]
\centering
\includegraphics[width=.36\textwidth]{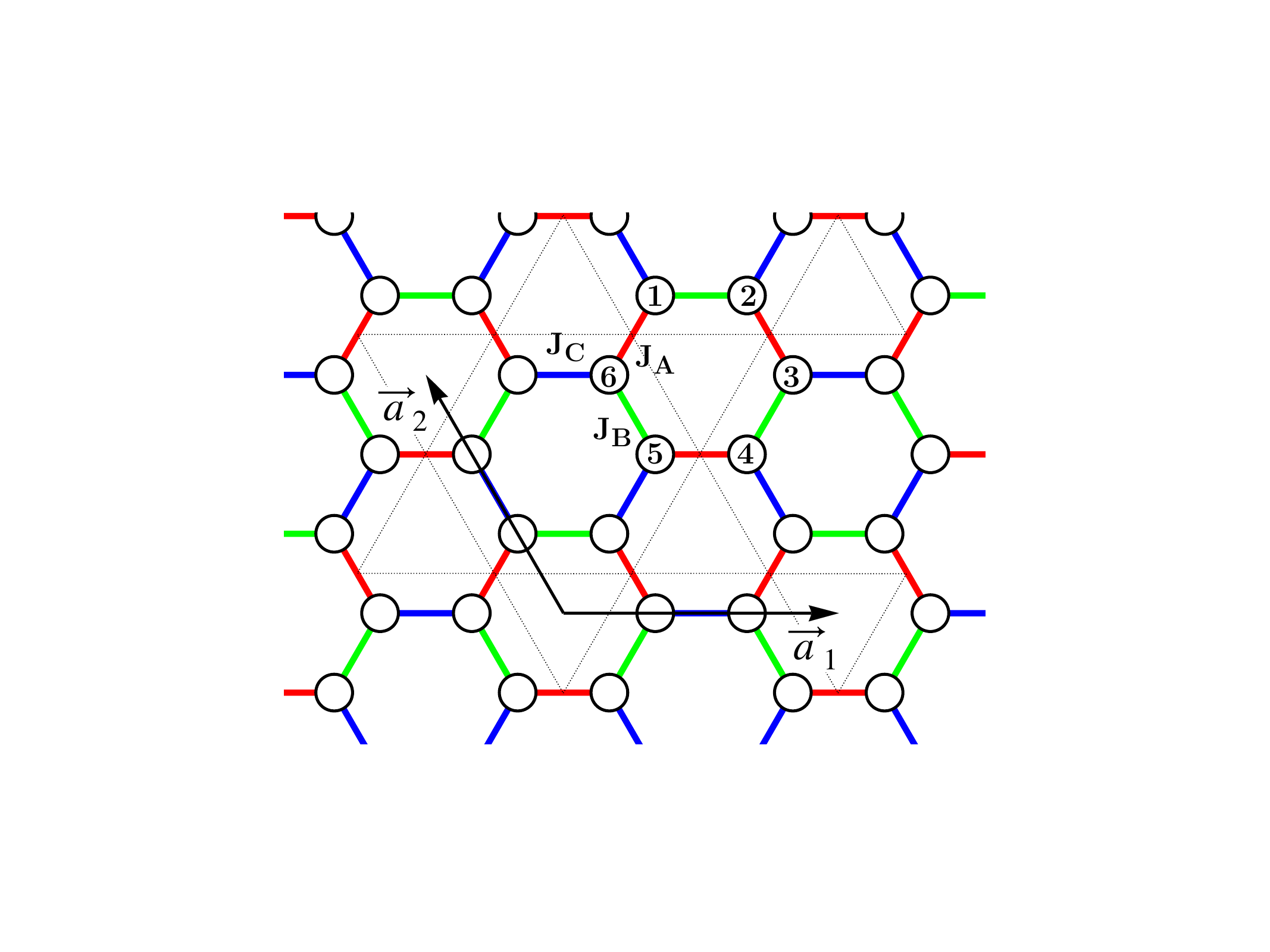}
\caption{The ABC Heisenberg model on kagome-honeycomb lattice. The exchange interactions $J_A$, $J_B$, and $J_C$ are antiferromagnetic, and they form a kagome superlattice (indicated by thin-dotted lines) on the honeycomb structure. The $\vec{a}_1$ and $\vec{a}_2$ are two primitive vectors of the underlying Bravais lattice.}
\label{fig:ABChoney}
\end{figure}

\begin{figure}[t] 
\centering
\includegraphics[width=.36\textwidth]{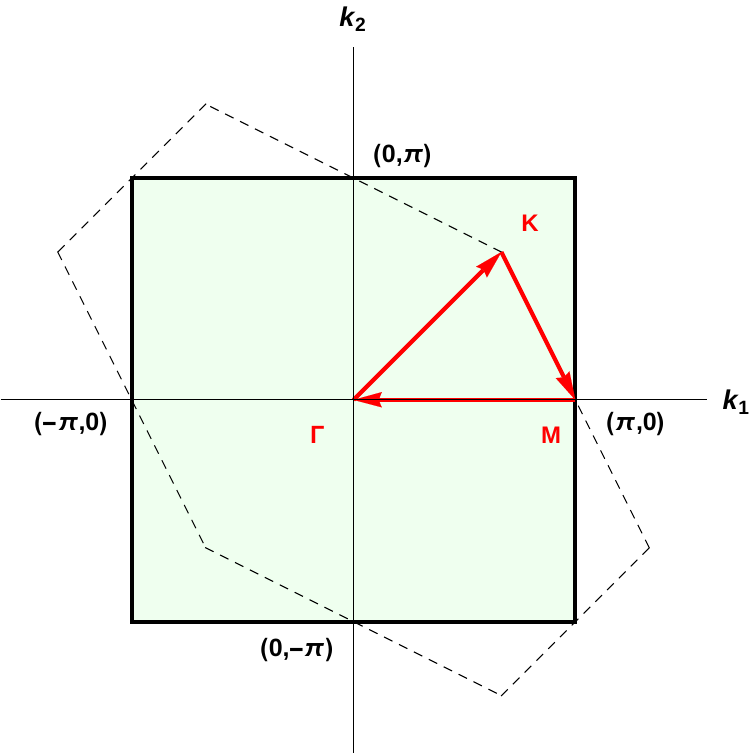}
\caption{The first Brillouin zone for the lattice in Fig.~\ref{fig:ABChoney}.  A point in this zone is the wavevector $\vec{k}=k_1\vec{b}_1+k_2\vec{b}_2$, where $\vec{b}_{1(2)}$ are reciprocal to $\vec{a}_{1(2)}$ such that $\vec{a}_i\cdot \vec{b}_j =\delta_{ij}$. The dashed line demarcates the Brillouin zone in the hexagonal form.}
\label{fig:BZ}
\end{figure}

The spin-1/2 Heisenberg model on kagome-honeycomb lattice is given by the Hamiltonian, 
$\Hhat_{ABC}  = \Hhat_A + \Hhat_B + \Hhat_C$, where 
\begin{subequations}
\label{eq:HABC}
\begin{align}
\Hhat_A & =  J_A\sum_{\vec{R}} \left(\vec{S}_{6\vec{R}}\cdot\vec{S}_{1\vec{R}}+\vec{S}_{2\vec{R}}\cdot \vec{S}_{3\vec{R}}+\vec{S}_{4\vec{R}} \cdot \vec{S}_{5\vec{R}}\right) \label{eq:HA} \\
\Hhat_B & =   J_B\sum_{\vec{R}} \left(\vec{S}_{1\vec{R}}\cdot\vec{S}_{2\vec{R}}+\vec{S}_{3\vec{R}}\cdot\vec{S}_{4\vec{R}}+\vec{S}_{5\vec{R}}\cdot\vec{S}_{6\vec{R}}\right) \label{eq:HB} \\
\Hhat_C &= J_C \sum_{\vec{R}} \left(\vec{S}_{1\vec{R}} \cdot \vec{S}_{4(\vec{R}+\vec{a}_2)}+\vec{S}_{3\vec{R}} \cdot \vec{S}_{6(\vec{R}+\vec{a}_1)} + \right. \nonumber \\ &   \left. { \hspace{1.6cm} } \vec{S}_{5\vec{R}} \cdot \vec{S}_{2(\vec{R}-\vec{a}_1-\vec{a}_2)}\right). \label{eq:HC}
\end{align}
\end{subequations}
Here, the exchange interactions $J_A$, $J_B$ and $J_C$ are all antiferromagnetic, and the lattice and the spin labels are as shown in Fig.~\ref{fig:ABChoney}. The basic structure is honeycomb, but the pattern of exchange interactions thereon is kagome. A primitive unit-cell of this so-called kagome-honeycomb lattice contains six spins marked here by the integers 1 to 6; $\vec{R}$ denotes the position of a primitive unit-cell. The vectors $\vec{a}_1=3a\xhat$ and $\vec{a}_2=3a(-\xhat+\sqrt{3}\yhat)/2$ are two primitive vectors of the underlying Bravais lattice. The corresponding Brillouin zone is shown in Fig.~\ref{fig:BZ}. We also call this model by a short name, the ABC model. 

Since the lattice in Fig.~\ref{fig:ABChoney} is bipartite, the antiferromagnetic ABC model on it is unfrustrated~\footnote{However, for mixed interactions, say with $J_A$ ferromagnetic and $J_{B(C)}$ antiferromagnetic, it is a frustrated model applicable to spin-1 kagome systems~Ref.~\cite{Hida2000,PG2018}}, and can be expected to realise N\'eel antiferromagnetic order. But the competition between the three exchange interactions, together with quantum fluctuations, provides enough scope for the spin-1/2 ABC model to destroy N\'eel order and realise a quantum paramagnetic ground state. Our goal is to study this competition.  Throughout this paper, the exchange interactions are taken to have values between $0$ and $1$ in such a way that $J_A+J_B+J_C=1$.

Since the uniform case, with $J_A$ $=$ $J_B$ $=$ $J_C$, is known to realise N\'eel order, even when the three exchange interactions are unequal, the N\'eel order is expected to survive in the vicinity of the point $(J_A,J_B,J_C)=(1/3,1/3,1/3)$ in the phase diagram. Far away from the uniform case, two interesting limiting cases arise. One in which only one type of bonds have non-zero exchange interaction, say $(J_A,J_B,J_C)=(1,0,0)$ , realises independent dimers.  The other case  in which only one type of bonds have zero exchange interaction, e.g. $(J_A,J_B,J_C)=(1-x,x,0)$ for $x\in[0,1]$, realises independent hexagons. In both the limiting cases, the ground state is a spin singlet and hence quantum paramagnetic (and in fact, spin liquid, as they break no symmetry of the model). 

\begin{figure}[t]
\centering
\includegraphics[width=0.46\textwidth]{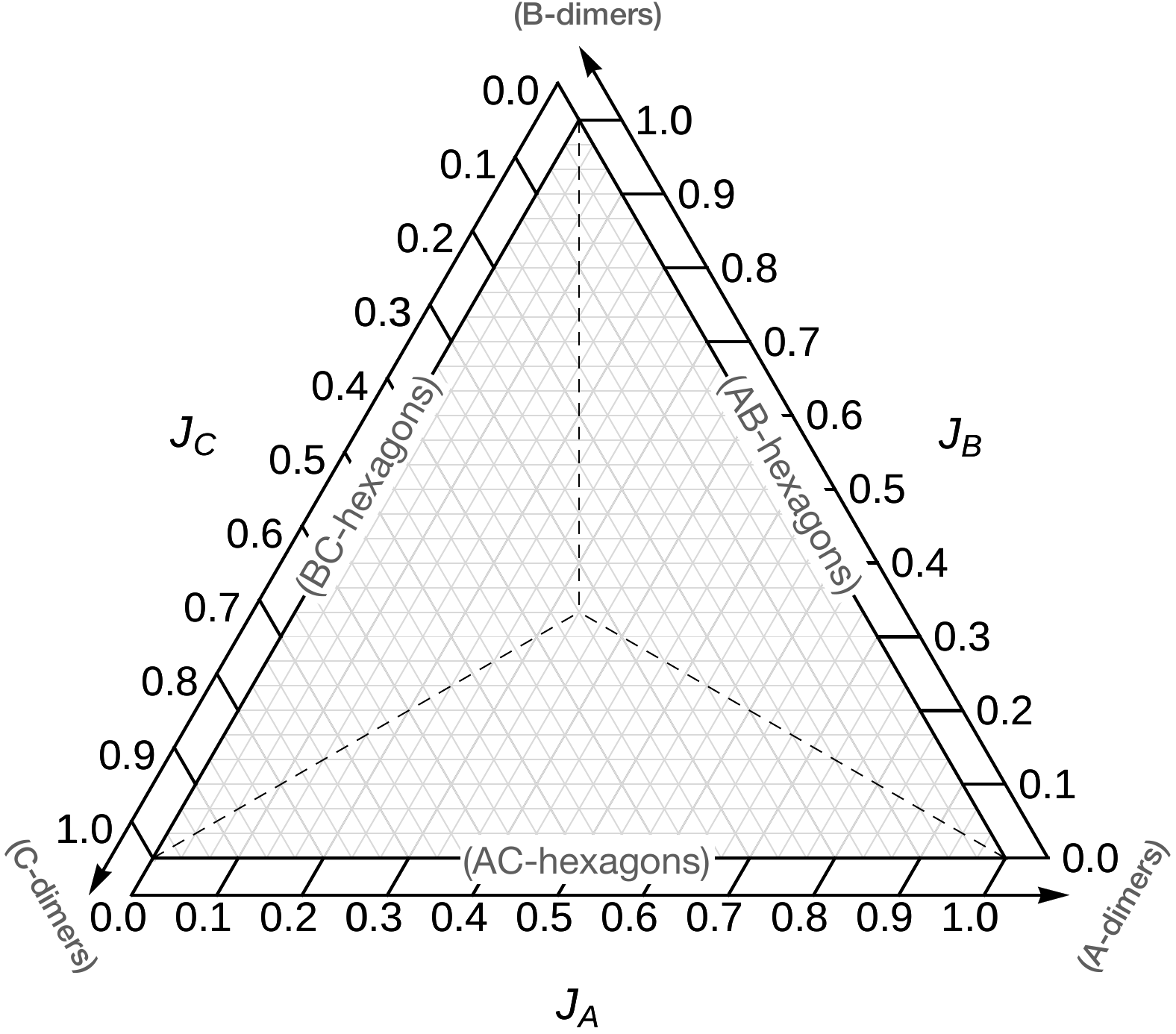}
\caption{Ternary representation of the space of exchange interactions  such that $J_A+J_B+J_C=1$. At the base of this equilateral triangle, $J_B=0$, which is the case of independent AC-hexagons. The apex (top corner) with $J_B=1$ corresponds to independent B-dimers. Likewise for the other two sides and corners. The centroid represents the uniform honeycomb antiferromagnet.} 
\label{fig:terBlank}
\end{figure}

In the ternary representation subject to the condition $J_A+J_B+J_C=1$, the space of interaction parameters is an equilateral triangle shown in Fig.~\ref{fig:terBlank}. The corners of this triangle correspond to independent dimers, and the sides to independent hexagons. The ground state of the ABC model is, therefore, bound to exhibit a quantum phase transition from the N\'eel antiferromagnetic phase in the interior around the centroid $(1/3,1/3,1/3)$ to a non-magnetic singlet phase outwards to the three sides of the ternary diagram. In the following sections, we make systematic analytical and numerical calculations to obtain the quantum phase diagram of the spin-1/2 ABC model on kagome-honeycomb lattice.

First we do the triplon analysis with respect to the hexagonal and the dimer singlet states. These are spin-gapped phases, for which the closure of the gap marks a quantum phase transition to the N\'eel phase. By following the triplon gap, and comparing the energies of these candidate states, we construct a quantum phase diagram. We then calculate spin stiffness and staggered magnetization by doing quantum Monte Carlo simulations. All these calculations produce a mutually agreeable phase diagram dominated by a quantum paramagnetic phase best described as a hexagonal singlet phase. 

\section{Triplon analysis}\label{sec:trip}
The basic framework of triplon analysis is to first identify such building-blocks of the system which in some limiting case realise singlet ground state locally independently, and then formulate an effective theory in terms of the low-energy triplet excitations of these building-blocks to describe the full system~\cite{Sachdev1990,Brijesh2010,PG2016,PG2018}. In this spirit, our ABC model can be viewed either as a system of coupled dimers, or coupled hexagons. For instance, we can consider the ABC model (see Fig.~\ref{fig:ABChoney}) as made of the A type bonds coupled by B and C bonds, or as made of the AB hexagons coupled via C bonds. (The other equivalent choices can be obtained by permuting A, B, C cyclically.). As mentioned earlier, in the limit of $J_B=J_C=0$, the model realises an exact dimer singlet (DS) ground state formed by the direct product of the singlets on the A bonds. When $J_C=0$, it similarly realises a hexagonal singlet (HS) ground state exactly. Thus, we have two ways of doing triplon analysis of the ABC model with respect to the two natural quantum paramagnetic states, DS or HS. Note that the exact DS state itself is a limit of an exact HS state, e.g. the $J_B=0$ case of the AB-hexagons is the same as having independent A-dimers. In pictorial terms (see Fig.~\ref{fig:terBlank}), the corners of the ternary diagram are the ends of its sides. It requires that we formulate the triplon analysis for the HS case in such a manner that, near the corners of the ternary diagram, it is consistent with the triplon analysis with respect to the DS state. Let us do it now, and see what we learn about the extent of the singlet phases as one moves inwards into the ternary diagram from its sides and corners. 
  
\subsection{Dimer singlet state \label{sec:DS}}
Assuming $J_A$ to be stronger than $J_B$ and $J_C$, we satisfy the Heisenberg interaction on the A-bonds exactly, and describe the spin operators in terms of the singlet and triplet eigenstates thereof. [The same is to be done with respect to B (or C) bonds, when $J_B$ (or $J_C$) is stronger than the rest.]. A convenient way to do this is to employ bond-operator representation, in which one uses bosonic operators for the singlet and triplets states of a bond~\cite{Sachdev1990,Brijesh2010}. It is simplified by treating the singlet bond-operator on every A-dimer as a mean singlet amplitude, $\bar{s}$, for the dimer singlet phase. The dynamics of the triplet excitations (triplons) in the DS phase is described using the triplet bond-operators.

The ABC model has three A-bonds per unit-cell (red bonds in Fig.~\ref{fig:ABChoney}). We label these bonds as $j=I, II, III$. Let $\alpha=x,y,z$ denote the three components of a spin. The six spins in a unit-cell at position $\vec{R}$ in the bond-operator representation (in a basic approximated form) can be written as:
\begin{subequations}
\label{eq:bo-DS}
\begin{eqnarray}
S^\alpha_{2\vec{R}} & \approx & \frac{\bar{s}}{2}(t^{\alpha\dag}_{\vec{R},I}+t^{\alpha}_{\vec{R},I}) \approx -S^\alpha_{3\vec{R}} \\
S^\alpha_{4\vec{R}} & \approx & \frac{\bar{s}}{2}(t^{\alpha\dag}_{\vec{R},II}+t^{\alpha}_{\vec{R},II}) \approx -S^\alpha_{5\vec{R}} \\
S^\alpha_{6\vec{R}} & \approx & \frac{\bar{s}}{2}(t^{\alpha\dag}_{\vec{R},III}+t^{\alpha}_{\vec{R},III}) \approx -S^\alpha_{1\vec{R}}
\end{eqnarray}
\end{subequations}
where $\that^\alpha_{\vec{R},j}$ and $\that^{\alpha\dag}_{\vec{R},j}$ are the triplet bond-operators. The bond-operators are also required to satisfy the constraint, $\bar{s}^2 + \sum_\alpha t^{\alpha\dag}_{\vec{R},j}t^{\alpha}_{\vec{R},j} =1$, to account for the physical dimension of the Hilbert space on every A-bond.  

Since the interaction on the A-bonds is treated exactly, we obtain the following expression for that part of the ABC model which comes from the A-bonds, i.e. the $\Hhat_A$ of Eq.~\eqref{eq:HA}, in terms of the singlet amplitude and the triplet bond-operators. 
\begin{eqnarray}
\hat{H}_{A} 
& = &  \sum_{\vec{R}}\sum_j \left(-\frac{3J_A}{4}\bar{s}^2 + \frac{J_A}{4} \sum_\alpha t^{\alpha\dag}_{\vec{R},j}t^{\alpha}_{\vec{R},j}\right)
\label{eq:hjahamilnew}
\end{eqnarray}
The triplets on different A-bonds interact and disperse on the lattice due to $\Hhat_B$ and $\Hhat_C$, i.e. Eqs.~\eqref{eq:HB} and \eqref{eq:HC}. We use Eqs.~\eqref{eq:bo-DS} to rewrite $\Hhat_B$ and $\Hhat_C$ in terms of the triplon operators. The constraint on the bond-operators is satisfied on average through a Lagrange multiplier $\lambda_0$ by adding the term, $\lambda_0 \sum_{\vec{R},j} \left(\bar{s}^2 + \sum_\alpha t^{\alpha\dag}_{\vec{R},j}t^{\alpha}_{\vec{R},j} - 1\right)$, to the triplon Hamiltonian. 

We find it convenient to write the triplon Hamiltonian  using canonical \textquotedblleft position\textquotedblright and ``momentum'' operators: $\hat{Q}^\alpha_{\vec{R},j}= \frac{1}{\sqrt{2}}(t^{\alpha\dag}_{\vec{R},j}+t^{\alpha}_{\vec{R},j})$ and $\hat{P}^\alpha_{\vec{R},j}=\frac{i}{\sqrt{2}}(t^{\alpha\dag}_{\vec{R},j}-t^{\alpha}_{\vec{R},j})$. They follow the relations $ \comm{\hat{Q}^\alpha_{\vec{R},j}}{\hat{P}^{\alpha^\prime}_{\vec{R^\prime},j^\prime}}=i\delta_{j,j^\prime}\delta_{\alpha,\alpha^\prime}\delta_{\vec{R},\vec{R}^\prime}$ and $  (\hat{P}^{\alpha}_{\vec{R},j})^2+(\hat{Q}^{\alpha}_{\vec{R},j})^2 = 2t^{\alpha\dag}_{\vec{R},j}t^{\alpha}_{\vec{R},j} + 1$. 
Their Fourier transformation is defined as: $ \hat{Q}^\alpha_{\vec{R},j} = \frac{1}{\sqrt{N_{uc}}}\sum_{\vec{k}}\hat{Q}^\alpha_{\vec{k},j}e^{i\vec{k}.\vec{R}}$ and $ \hat{P}^\alpha_{\vec{R},j} = \frac{1}{\sqrt{N_{uc}}}\sum_{\vec{k}}\hat{P}^\alpha_{\vec{k},j}e^{i\vec{k}.\vec{R}}$, where $N_{uc}$ is the total number of unit-cells, and the wavector $\vec{k}$ lies in the Brillouin zone drawn in Fig.~\ref{fig:BZ}. 
Moreover, $(\hat{Q}^{\alpha}_{\vec{k},j} )^\dag = \hat{Q}^{\alpha}_{-\vec{k},j}$, 
and likewise for $\hat{P}^{\alpha}_{\vec{k},j}$. 

We obtain the following effective Hamiltonian for the triplon dynamics with respect to the DS state. 
\begin{equation}
 \hat{H}_{tDS}=\epsilon N_{uc} + \frac{1}{2}\sum_{\vec{k},\alpha} \left\{ \lambda \bold{P}^{\alpha^\dag}_{\vec{k}}\mathbb{I}_{3}\,\bold{P}^{\alpha}_{\vec{k}} + \bold{Q}^{\alpha^\dag}_{\vec{k}} V_{\vec{k}}\, \bold{Q}^{\alpha}_{\vec{k}} \right\}
 \label{eq:totalhamilf}
\end{equation}
Here, $\lambda=\lambda_0 + \frac{J_A}{4}$,  $\epsilon=3\bar{s}^2\lambda + \frac{3J_A}{4}-\frac{15}{2}\lambda-3J_A\bar{s}^2$, and $\mathbb{I}_{3}$ is the $3\cross3$ identity matrix; $\bold{P}^{\alpha}_{\vec{k}}$, $\bold{Q}^{\alpha}_{\vec{k}}$ and $V_{\vec{k}}$ are given below. 
\begin{equation}
\bold{P}^{\alpha}_{\vec{k}}
=
\begin{pmatrix}
    P^{\alpha}_{\vec{k},\RomanNumeralCaps{1}} \\
    P^{\alpha}_{\vec{k},\RomanNumeralCaps{2}} \\
    P^{\alpha}_{\vec{k},\RomanNumeralCaps{3}} 
\end{pmatrix},
\bold{Q}^{\alpha}_{\vec{k}}
=
\begin{pmatrix}
    Q^{\alpha}_{\vec{k},\RomanNumeralCaps{1}} \\
    Q^{\alpha}_{\vec{k},\RomanNumeralCaps{2}} \\
    Q^{\alpha}_{\vec{k},\RomanNumeralCaps{3}} 
\end{pmatrix}
\end{equation}

\begin{equation}
\begin{aligned}
 V_{\vec{k}} = & ~ {} \lambda \mathbb{I}_{3} - \frac{J_B\bar{s}^2}{2}
 \begin{pmatrix}
    0 && 1 && 1 \\
    1 && 0 && 1 \\
    1 && 1 && 0 
 \end{pmatrix}
 \\
& - \frac{J_C\bar{s}^2}{2}
 \begin{pmatrix}
    0 && e^{i k_3} && e^{i k_1} \\
    e^{-i k_3} && 0 && e^{-i k_2} \\
    e^{-i k_1} && e^{i k_2} && 0 
 \end{pmatrix}
 \end{aligned}
 \label{eq:nuk}
\end{equation}
Note that  $k_1=\vec{k}.\vec{a}_1$, $k_2=\vec{k}.\vec{a}_2$ and $k_3 =k_1+k_2$. The eigenvalues of $V_{\vec{k}}$ are found to be 
\begin{equation}
\omega_{\vec{k},j} = \sqrt{\lambda( \lambda-2\bar{s}^2\zeta_{\vec{k},j} )}
\end{equation}
where $\zeta_{\vec{k},I} = -\frac{1}{4}(J_B + J_C)$ is $\vec{k}$ independent, while
\[\zeta_{\vec{k}, II} = \frac{1}{8}\left[J_B + J_C + \sqrt{9J_B^2-6J_BJ_C+9J_C^2+8J_BJ_Cf^0_{\vec{k}}}\,\right] \] and
\[ \zeta_{\vec{k}, III} = \frac{1}{8}\left[J_B + J_C - \sqrt{9J_B^2-6J_BJ_C+9J_C^2+8J_BJ_Cf^0_{\vec{k}}}\, \right] \] depend on $\vec{k}$ through $f^0_{\vec{k}}  = \cos{k_1}+\cos{k_2}+\cos{k_3}$. Knowing these $\omega_{\vec{k},j}$'s (the triplon dispersions of $\Hhat_{tDS}$) gives the following ground state energy per unit-cell. 
\begin{equation}
 E_{gDS} = \epsilon + \frac{3}{2N_{uc}}\sum_{\vec{k}}\sum_j\omega_{\vec{k},j}
 \label{eq:Eg}
\end{equation}
Minimizing the $E_{gDS}$ with respect to  $\bar{s}^2$ and $\lambda$ leads to the following equations,
\begin{subequations}
\label{eq:sceDS}
\begin{align}
 \lambda &= J_A + \frac{\lambda}{2N_{uc}}\sum_{\vec{k},j} \frac{\zeta_{\vec{k},j}}{\omega_{\vec{k},j}} \\
 \bar{s}^2 &= \frac{5}{2} - \frac{1}{2 N_{uc}}\sum_{\vec{k},j}\frac{\lambda-\bar{s}^2\zeta_{\vec{k},j}}{\omega_{\vec{k},j}}
\end{align}
\end{subequations}
whose self consistent solution determines the dimer singlet phase for the ABC model. 

Before solving these equations for $\lambda$ and $\sbar^2$, let us also formulate a theory of triplon dynamics with respect to the hexagonal singlet state. Then, we will present and discuss their findings together. 

\subsection{Hexagonal singlet state \label{sec:HS}}  
When $J_C=0$, the ABC model is a collection of independent AB-hexagons (see Figs.~\ref{fig:ABChoney} and \ref{fig:terBlank}). So, when $J_C$ is non-zero (but somewhat weaker than $J_A$ and $J_B$), it is reasonable to formulate a theory of the ABC model 
in terms of the eigenstates of the AB-hexagons. In doing so, we satisfy two interactions ($J_A$ and $J_B$) exactly, which certainly makes for a better case (than the dimer case of the previous section, where only one interaction, $J_A$, was exactly satisfied). 

\begin{figure}[t]
  	\centering
  	\includegraphics[width=0.48\textwidth]{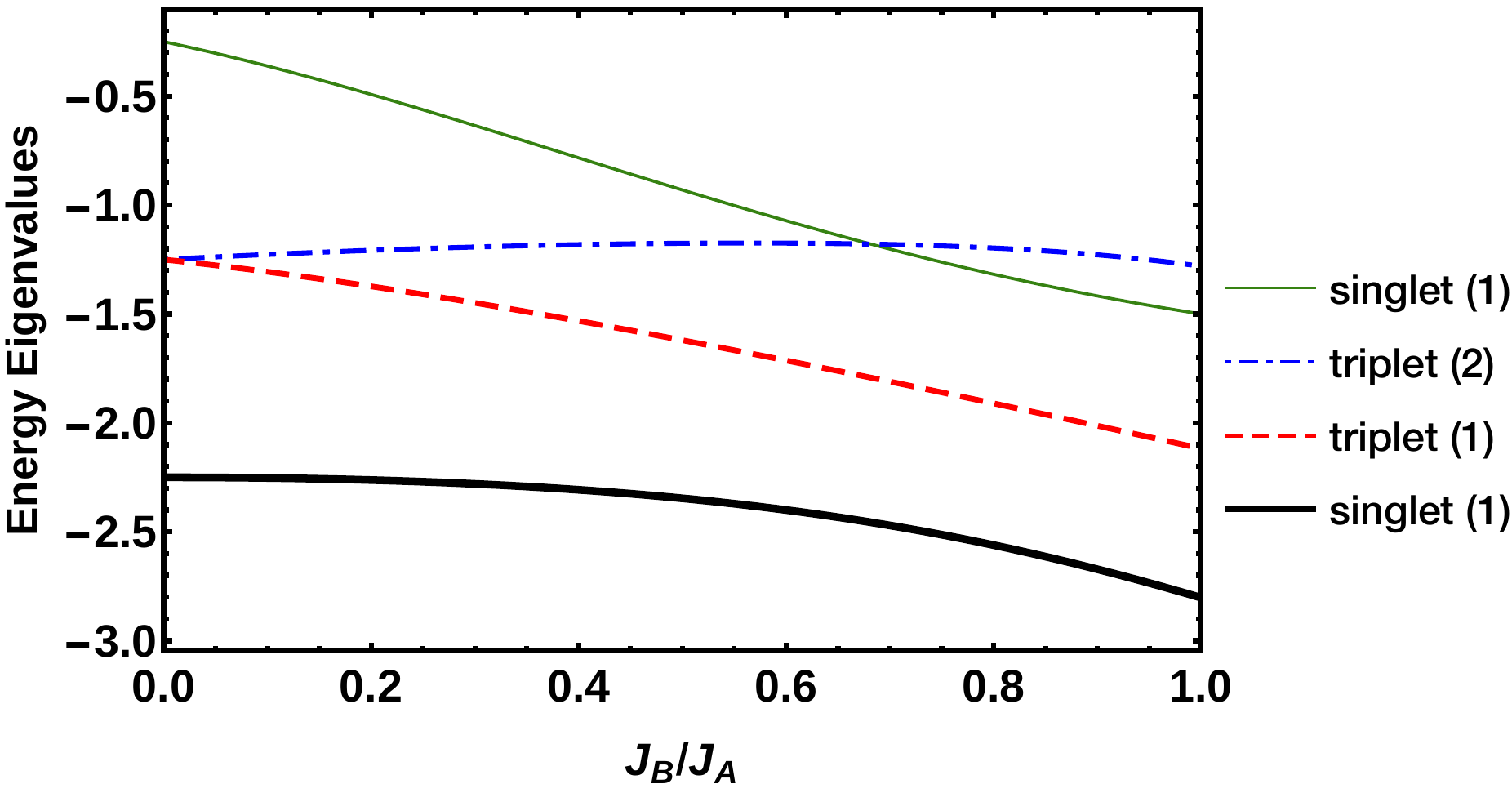}
  	\caption{Low-energy spectrum of a spin-1/2 hexagon with alternating nearest-neighbour exchange interactions $J_A$ and $J_B$; refer to Appendix~\ref{app:eigHex}. The lowest eigenvalue (thick black line) corresponds to a unique singlet, $\ket{s}$, and the second lowest (dashed red) to a triplet, $\ket{t_{m0}}$. Then, there are two degenerate triplets (dot-dashed blue, $\ket{t_{m1}}$ and $\ket{t_{m\bar{1}}}$) crossed by another singlet (thin green). The other higher energy states, not relevant for the triplon analysis in Sec.~\ref{sec:HS}, are not shown. The ABC Heisenberg model of Eq.~\eqref{eq:HABC} is a system of such AB-hexagons coupled via the exchange interaction $J_C$.}
  	\label{fig:eig-ABhex}
  \end{figure}

The exact eigenspectrum of the Heisenberg model of a single AB-hexagon is evaluated in Appendix~\ref{app:eigHex}, of which the lowest few eigenstates are plotted in Fig.~\ref{fig:eig-ABhex}. Here, the ground state is a unique singlet, separated from the first excited state (which is a triplet) by a finite energy. When these hexagons are coupled via $J_C$, one would expect the ground state of the full model to be a hexagonal singlet (HS) state renormalised by triplet fluctuations, but protected by triplon gap. For sufficiently strong $J_C$, either this triplon gap will close causing a phase transition to an ordered antiferromagnetic phase, or another state may level-cross. What one minimally needs to carry out such an anaylsis is the lowest singlet and triplet eigenstates. But as noted earlier, the triplon analysis based on hexagonal states is desired to be such that its approach to the dimer limit (for small $J_B$ or $J_A$) is appropriate. Figure~\ref{fig:eig-ABhex} suggests that we should take into consideration the next two degenerate triplets also, because these two become degenerate with the lowest triplet (as for three independent dimers) when $J_B$ tends to zero. Taking three triplets considerably enhances the complexity of the triplon analysis, but it does give us a theory that works very well.

These eigenstates are identified by their total spin and two other quantum numbers, $m$ and $\nu$ corresponding respectively to the $z$-component of the total spin and the threefold rotation of the hexagon. The $\nu$ takes values $0,1$, $\bar{1} (=-1)$ (for the rotation eigenvalues $1,\omega,\omega^2$, respectively), and $m$ takes values $0,\pm1,\pm2$, $\pm3$. Refer to Appendix~\ref{app:eigHex} for more details. Of the states presented in Fig.~\ref{fig:eig-ABhex}, we denote the singlet ground state as $|s\rangle$ and its energy as $E_{s}$; it belongs to $m=0,\nu=0$ subspace. The triplets are denoted as $|t_{m\nu}\rangle$ with $m=0,\pm1$. The lowest energy triplet corresponds to $\nu=0$ with energy denoted as $E_{t0}$, and the next two triplets correspond to $\nu=1,\bar{1}$ with energy $E_{t1}$. Note that for $J_B/J_A \gtrsim 0.68$, a higher energy singlet becomes slightly lower in energy than $E_{t1}$. But unlike the triplets, this second singlet makes no direct matrix elements (of the spin operators) with the singlet ground state. So, its effect on a low-energy theory based on the hexagonal states 
is negligible; we have checked this. Hence, we consider $|s\rangle$ and $|t_{m\nu}\rangle$ only (a total 10 states per hexagon) to formulate a theory with respect to the hexagonal singlet ground state. 

Like the bond-operators employed for the dimer case, we now introduce the bosonic operators, 
$\shat_{\vec{R}}$ and $\that_{m\nu,\vec{R}}$, 
corresponding to the hexagonal singlet and triplet states at position $\vec{R}$~\cite{PG2018}. Next we replace the singlet operator on every hexagon by a mean amplitude $\bar{s}$ that accounts for the hexagonal singlet background. Then, we write the six spins (labelled as $l=1$ to $6$) on an AB-hexagon in terms of the triplet operators as follows. 
\begin{subequations}
\label{eq:repHex}
\begin{align}
		S_{l\vec{R}} ^z\approx&~\bar{s}\left[\mathcal{C}^l_{00} (\that_{00,\vec{R}}+\that_{00,\vec{R}}^\dag)+ (\mathcal{C}^l_{01}\that_{01,\vec{R}}+\mathcal{C}^{l*}_{01}\that_{0\bar{1},\vec{R}} + {\rm h.c})\right]\\
		S_{l\vec{R}}^+\approx&~\bar{s} \left[ \mathcal{C}^l_{\bar{1}0} (\that_{\bar{1}0,\vec{R}}-\that_{10,\vec{R}}^\dag)+\mathcal{C}^l_{\bar{1}1}(\that_{\bar{1}1,\vec{R}}-\that^\dag_{1\bar{1},\vec{R}}) \right. \nonumber \\
		& \left. ~~~ +\mathcal{C}^{l*}_{\bar{1}1}(\that_{\bar{1}\bar{1},\vec{R}}-\that_{11,\vec{R}}^\dag) \right]
	\label{eq:HSre}
\end{align}
\end{subequations}
Here, the coefficients $\mathcal{C}^l_{00}$, $\mathcal{C}^l_{01}$ et cetera are the matrix elements between the singlet and the triplet states. Refer to Appendix~\ref{app:eigHex} for more details on this representation. The constraint in this case is $ \sbar^2 +\sum_{m\nu} t^{\dag}_{m\nu,\vec{R}}t_{m\nu,\vec{R}} = 1$. 

The $\Hhat_A +\Hhat_B$ part of the ABC model in this representation reads as:
\begin{equation}
\Hhat_A + \Hhat_B \approx E_s \bar{s}^2 N_{uc} + \sum_{\vec{R}, m\nu} E_{m\nu}\that^{\dag}_{m\nu,\vec{R}} \that_{m\nu,\vec{R}}
\end{equation}
where $E_{m0}=E_{t0}$ and $E_{m1}=E_{m\bar{1}}=E_{t1}$. The interaction between the AB-hexagons comes from $\Hhat_C$, which is now re-expressed using the representation in Eqs.~\eqref{eq:repHex}. Moreover, the constraint is taken into account by adding the term $\lambda \sum_{\vec{R}}(\bar{s}^2 + \sum_{m\nu} t^{\dag}_{m\nu,\vec{R}}t_{m\nu,\vec{R}} -1)$ to the Hamiltonian through a Lagrange multiplier $\lambda$. By Fourier transforming the triplon operators as, $\that_{m\nu,\vec{R}} = \frac{1}{\sqrt{N_{uc}}} \sum_{\vec{k}}e^{i \vec{k}\cdot\vec{R}}\that_{m\nu,\vec{k}}$, we finally get the following triplon Hamiltonian for the hexagonal singlet case. 
\begin{align}
		\Hhat_{tHS} = \epsilon_0N_{uc} + \sum_{\vec{k}} \Psi_{\vec{k}}^\dag \mathcal{H}^{ }_{\vec{k}}\Psi^{ }_{\vec{k}}
\label{eq:tHS}
\end{align}
Here, $\epsilon_0=E_s\bar{s}^2 + \lambda\bar{s}^2 -\frac{11\lambda}{2} - \frac{3}{2}(E_{t0} + 2E_{t1})$, $\mathcal{H}_{\vec{k}}$ is an $18\cross18$ matrix in the Nambu basis given in Appendix~\ref{app:HShamil}, and $\Psi^\dag_{\vec{k}}$ is the following row vector of triplon creation and annihilation operators; $\Psi_{\vec{k}}$ is its Hermitian conjugate. 
\begin{equation}
\Psi^\dag_{\vec{k}} = \begin{pmatrix}
\that^\dag_{00,\vec{k}} & \that^\dag_{10,\vec{k}} & \that^\dag_{\bar{1}0,\vec{k}} & \cdots & \that_{0\bar{1},\vec{-k}} & \that_{1\bar{1},\vec{-k}} & \that_{\bar{1}\bar{1},\vec{-k}}
	\end{pmatrix}
	\label{eq:numbu}
\end{equation}

We diagonalize $\Hhat_{tHS}$ using Bogloliubov transformation, and obtain nine triplon dispersions, $2\epsilon_{i\vec{k}}$, in terms of which the ground state energy can be written as:
\begin{align}
E^{ }_{gHS} &= \epsilon_0 + \frac{1}{N_{uc}}\sum_{\vec{k}}\sum_{i=1}^{9} \epsilon_{i\vec{k}}
	\label{eq:HSEg}
\end{align}
The following self-consistent equations for $\bar{s}$ and $\lambda$ are obtained by minimizing $E_g$, i.e. $\frac{\partial E_g}{\partial \bar{s}}=0$ and $\frac{\partial E_g}{\partial \lambda}=0$.

\begin{subequations}
\label{eq:sceHS}
\begin{align}
\lambda &= -E_{s} - \frac{1}{N_{uc}} \sum_{\vec{k}} \sum_{i=1}^9 \frac{\partial \epsilon_{i\vec{k}}}{\partial \bar{s}^2}\\
\bar{s}^2 &= \frac{11}{2} - \frac{1}{N_{uc}}\sum_{\vec{k}}\sum_{i=1}^{9} \frac{\partial \epsilon_{i\vec{k}}}{\partial \lambda}
\end{align}
\end{subequations}

\subsection{Quantum phase diagram from triplon analysis}
\begin{figure}[t]
 \centering
 \includegraphics[width=0.48\textwidth]{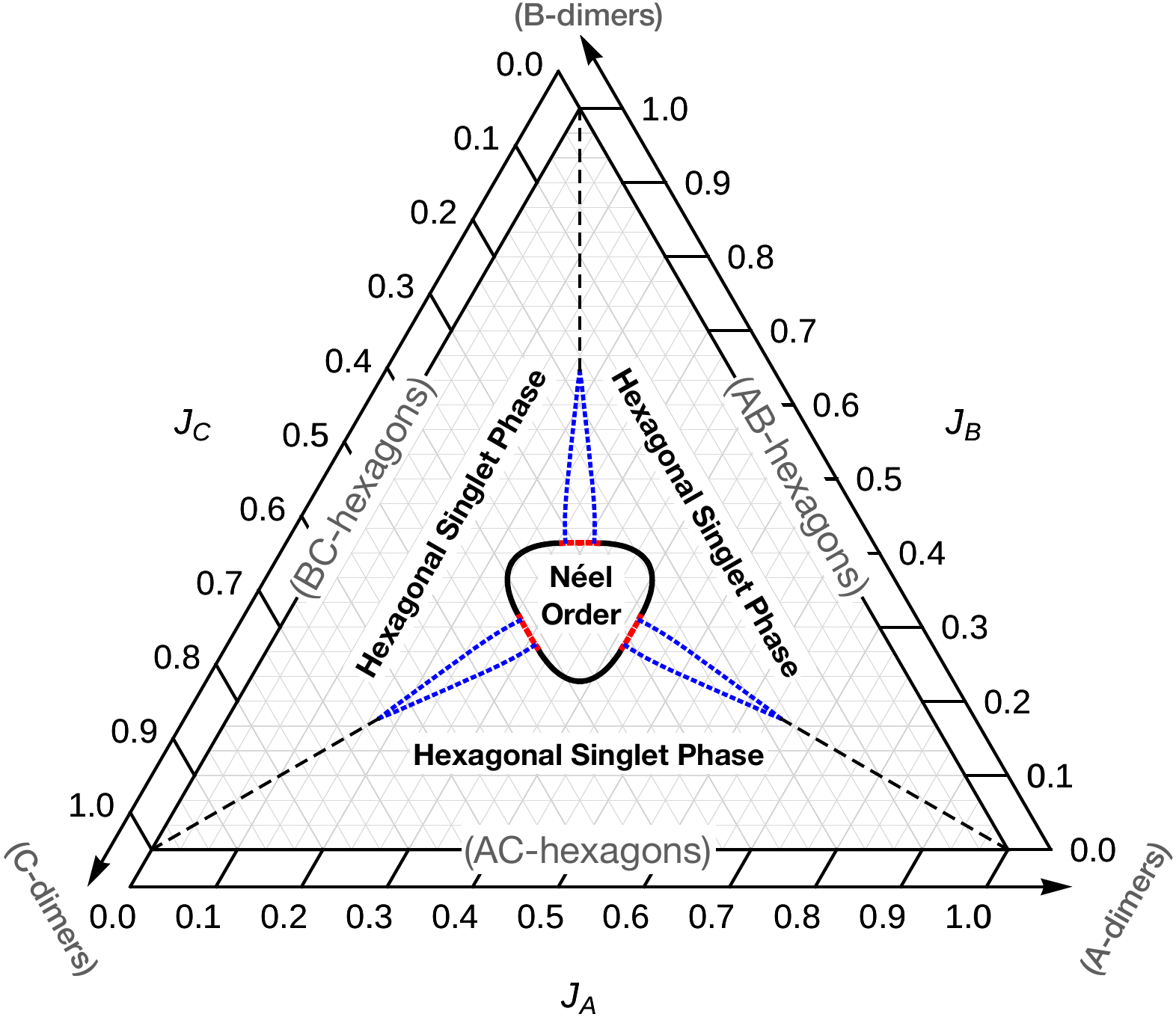}
 \caption{Quantum phase diagram of the ABC Heisenberg antiferromagnet on kagome-honeycomb lattice from triplon analysis. It is dominated by the spin-gapped hexagonal singlet phase on the three sides, with a small region of N\'eel antiferromagnetic phase in the middle, and small competing regions at the interfaces between the hexagonal singlet phases.}
 \label{fig:QPD-trip}
 \end{figure}

We solve Eqs.~\eqref{eq:sceDS} and Eqs.~\eqref{eq:sceHS} numerically. It gives us the triplon dispersions and the ground state energy with respect to the DS and HS states, respectively. By comparing their energies, and by following the triplon gap, we obtain a quantum phase diagram presented in Fig.~\ref{fig:QPD-trip}. As anticipated, it has in the middle a small region of N\'eel antiferromagnetic phase, which is surrounded on all three sides 
by a quantum paramagnetic phase described pretty well for the most part as an hexagonal singlet phase (with respect to the AB, BC or AC hexagons in the three triangular parts of the ternary diagram).
 
 \begin{figure}[t]
\centering
\includegraphics[width=0.46\textwidth]{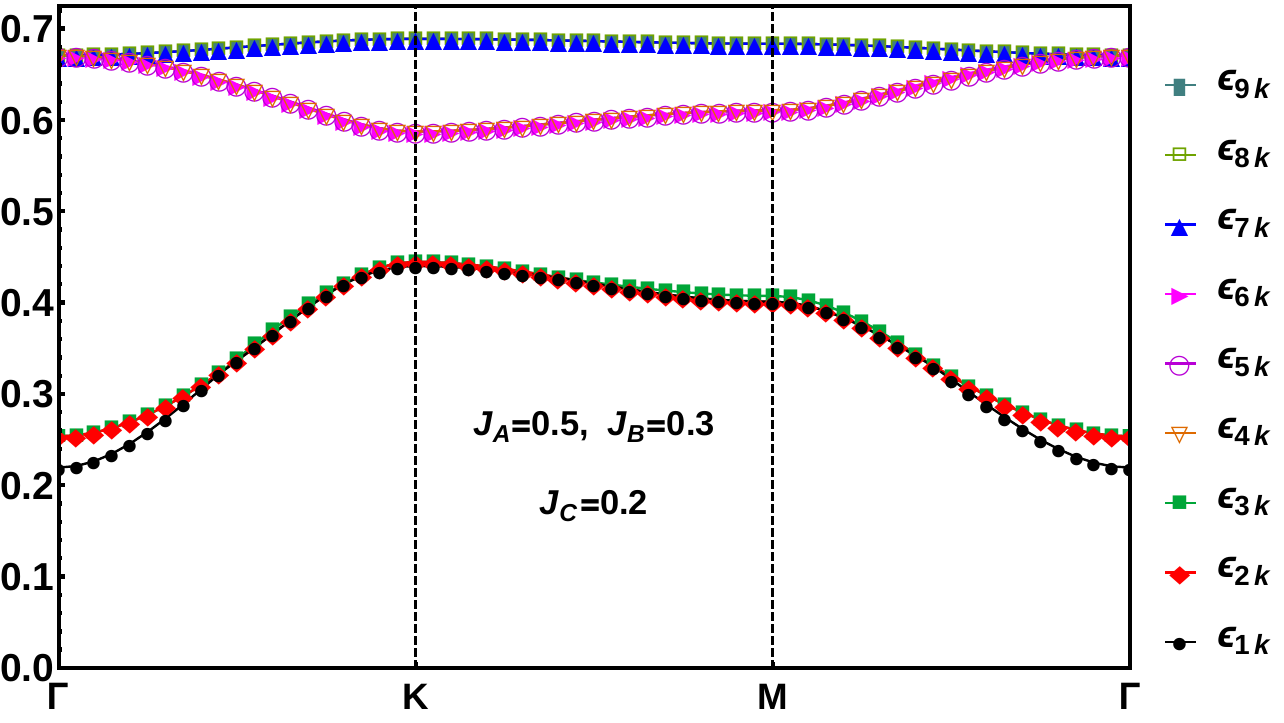}
\caption{Triplon dispersions in the gapped HS phase along the symmetry directions shown in Fig.~\ref{fig:BZ}. The triplon energy gap comes from the $\Gamma$ point.}
\label{fig:dis-HS}
\end{figure}

\begin{figure}[t]
 \centering
 \includegraphics[width=0.4\textwidth]{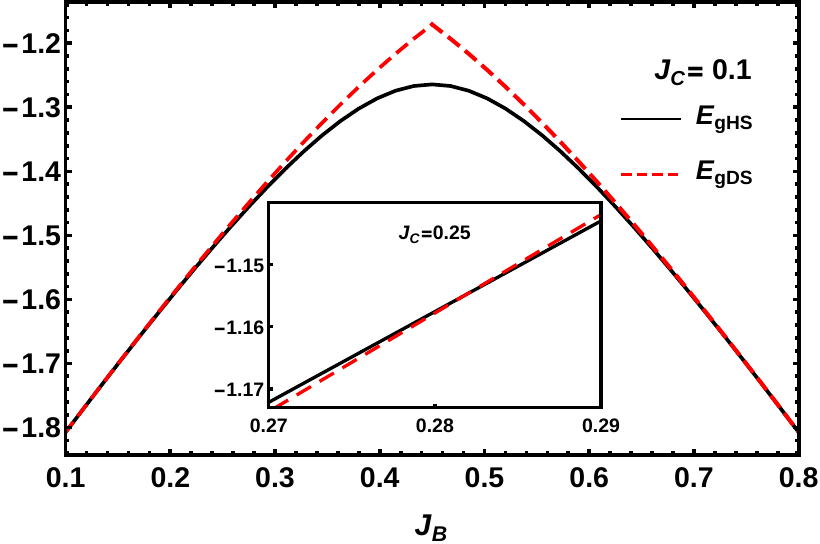}
 \caption{Ground state energies of the hexagonal singlet ($E_{gHS}$) and the dimer singlet  ($E_{gDS}$) phases from triplon analysis for $J_C=0.1$. Inset shows a level-crossing between the DS and the HS states by varying $J_B$ for $J_C=0.25$. Such level-crossings occur across the blue dotted lines in Fig.~\ref{fig:QPD-trip}.}
 \label{fig:Eg-trip}
 \end{figure}

For the concreteness of discussion, let us focus in Fig.~\ref{fig:QPD-trip} on the triangular region on right-hand-side, given by $0 \le J_C \le 1/3$ $\bigcap$ $ J_A \ge J_C$  $\bigcap$ $ J_B \ge J_C $. It is formed by joining the top corner, right corner and the centroid. (The other two similar regions are related to this one by the cyclic permutation of $J_A$, $J_B$, $J_C$.). In this region of the phase diagram, for $J_C=0$, we have independent AB-hexagons with exact HS ground state having a finite energy gap to triplet excitations. We find that for small non-zero $J_C$, the triplon excitations of the renormalized HS state are still gapped, and the mean singlet weight per hexagon, $\sbar^2$, is close to 1. See Fig.~\ref{fig:dis-HS} for triplon dispersions in the gapped HS phase. We also find that closer to the corners of the ternary diagram, the results from the HS state triplon analysis correctly approach the DS case. See Fig.~\ref{fig:Eg-trip} for the energies of the HS and DS states from triplon analysis as a function of $J_B$ for $J_C=0.1$. For small values of the inter-hexagon interaction (i.e., $J_C$ here), the HS state is always lower in energy than the DS state. Hence, the model clearly realises the HS phase near the three sides of the ternary diagram. This behaviour from the AB-hexagon side continues upto $J_C \approx 0.18$.

 \begin{figure}[t]
\centering
\includegraphics[width=0.43\textwidth]{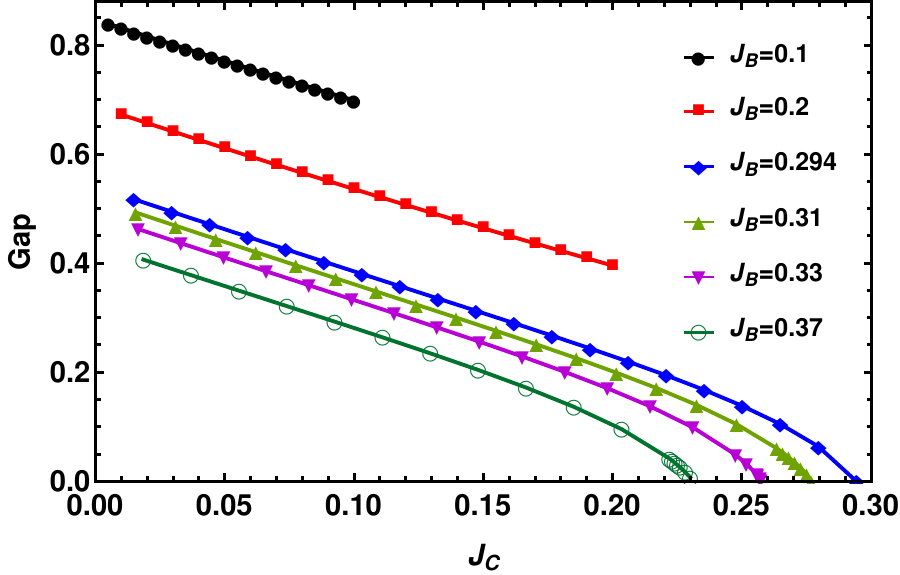} 
\caption{Triplon gap of the AB-hexagonal singlet state along fixed $J_B$ lines for $J_C \in [0,J_B]$.}
\label{fig:gapHS}
\end{figure}

For $J_C \ge 0.18$, the DS state is found to become lower in energy than the HS state, but only when either $J_A $ or $J_B$ is very close to $J_C$. This level-crossing happens across the blue-dotted lines in Fig.~\ref{fig:QPD-trip}; also see the inset of Fig.~\ref{fig:Eg-trip}. The gapped HS phase with respect to the AB-hexagons still holds good for the most part, except very close to the interface with AC (or BC) hexagonal phase. At the interface between, say, the AB and AC hexagonal phases, the B and C bonds would naturally compete to partner with the A bonds to form the respective HS state. So, when the exchange interactions of comparable values on B and C bonds are strong enough, it is possible that it is favourable for neither of them to partner with A. This is what this level-crossing seems to be hinting at. In the present analysis, the DS state of A-dimers happens to offer an alternative for the B and C bonds to be treated freely and not bound to A~\footnote{In Ref.~\cite{BKThesis}, the model with exact dimer singlet ground state on kagome-honeycomb lattice has $(J_A,J_B,J_C)$ $=$ $(0.5,0.25,0.25)$, apart from some frustrating interactions. Notably, it lies precisely on the interface inside the DS phase marked by the blue-dotted lines in Fig.~\ref{fig:QPD-trip} here.}. But it does not exclude the possibility of an alternate description of this competing cross-over region.

At $J_C = 0.23$, we for the first time find the HS phase to become gapless along the $J_A=J_B$ line. This closing of the triplon gap (at the $\Gamma$ point in the Brillouin zone) is found to occur in a continuous manner. See Fig.~\ref{fig:gapHS} for the triplon gap in the HS phase. For $J_C>0.23$, we get a finite region of the gapless HS phase in the middle. It is a common knowledge that the gapless triplons describe magnetic order~\cite{Sachdev1990,PG2016}. Hence, what we find here is a quantum phase transition from the gapped hexagonal singlet phase to the N\'eel antiferromagnetic phase. The thick black line in Fig.~\ref{fig:QPD-trip} is the boundary of this quantum phase transition.

Upon increasing the $J_C$ further, there comes a stage at $J_C\sim 0.27$, when the gapped HS phase is lost. Now the competing region described here as a gapped DS phase is found to be directly crossed by the N\'eel state (e.g., at $J_A=0.41$ along $J_B=J_C$ line). This level-crossing (shown by the red dashed lines in Fig.~\ref{fig:QPD-trip}) is obtained by comparing the energy of the DS (and the HS) state with that of the N\'eel state from spin-wave theory; see Appendix~\ref{app:SW} for spin-wave calculation. It ought be pointed out here that, pretty much where the DS state is crossed by the N\'eel state, the HS state (although energetically slightly ill-favoured here) still exhibits a continuous phase transition to the Neel phase. These small competing regions appear to be more complex.

\section{Quantum Monte Carlo Simulation}\label{sec:num}
In order to challenge and confirm the quantum phase diagram obtained from triplon analysis, we also employ quantum Monte Carlo method to study this problem. We are able to do so because our ABC Heisenberg model on kagome-honeycomb lattice is bipartite and un-frustrated, and hence amenable to QMC approach. We use the well-known stochastic series expansion (SSE) formulation of QMC~\cite{Sandvik1997,Sandvik2002}, which is exact albeit stochastic. Within this framework, the physical quantities such as the staggered magnetization, $m_s$, and the spin stiffness, $D_s$, can be calculated. The latter is defined as $D_s = \frac{1}{N} \frac{\partial^2 F (\phi)}{\partial \phi^2}$, where $F$ is the free energy of the system, $N$ is the total number of spins (sites) of the honeycomb lattice, 
and $\phi$ is the twist angle imposed on the periodic boundary condition. This quantity is considered to be a clean marker of the transition from an ordered ($D_s \neq 0$) to disordered phase ($D_s=0$). Within the SSE simulations, the spin stiffness is extracted using the winding number fluctuations as established in [\onlinecite{Sandvik1997}]. 
The former quantity, defined as $m_s = \langle {\bf m}_s\rangle = \frac{1}{N} \sum_{i} \langle S_{i,u}^z - S_{i,v}^z \rangle$, is the order parameter of the N\'eel phase. Here, $i$ is summed over the two-site unit-cells of the honeycomb lattice, and $u$ and $v$ denote the two sublattices. In the QMC simulations for finite size systems, what we calculate is the average value, $\langle {\bf m}_s^2\rangle$, which in the thermodynamic limit gives the square of the N\'eel order parameter (i.e., $m_s^2$). 

\begin{figure}[t]
\centering
\includegraphics[width=0.48\textwidth]{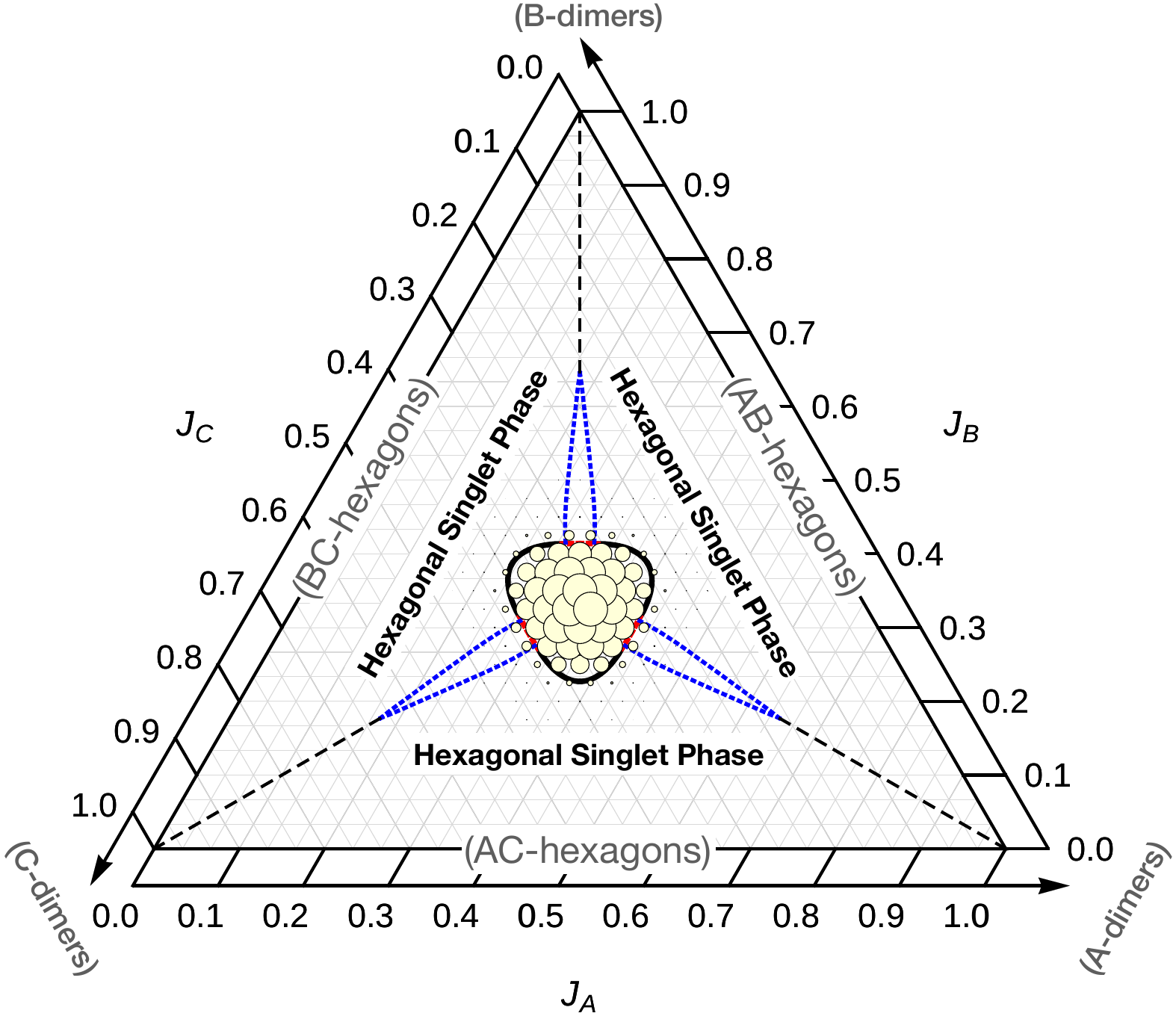}
\caption{The spin-stiffness data (pale yellow circles) from QMC calculations plotted together with the quantum phase diagram from triplon analysis (Fig.~\ref{fig:QPD-trip}). The radii of the circles indicate the strength of the N\'eel order.}
\label{fig:QPD-trip-stiff}
\end{figure}

\begin{figure}[t]
\centering
\includegraphics[width=0.46\textwidth]{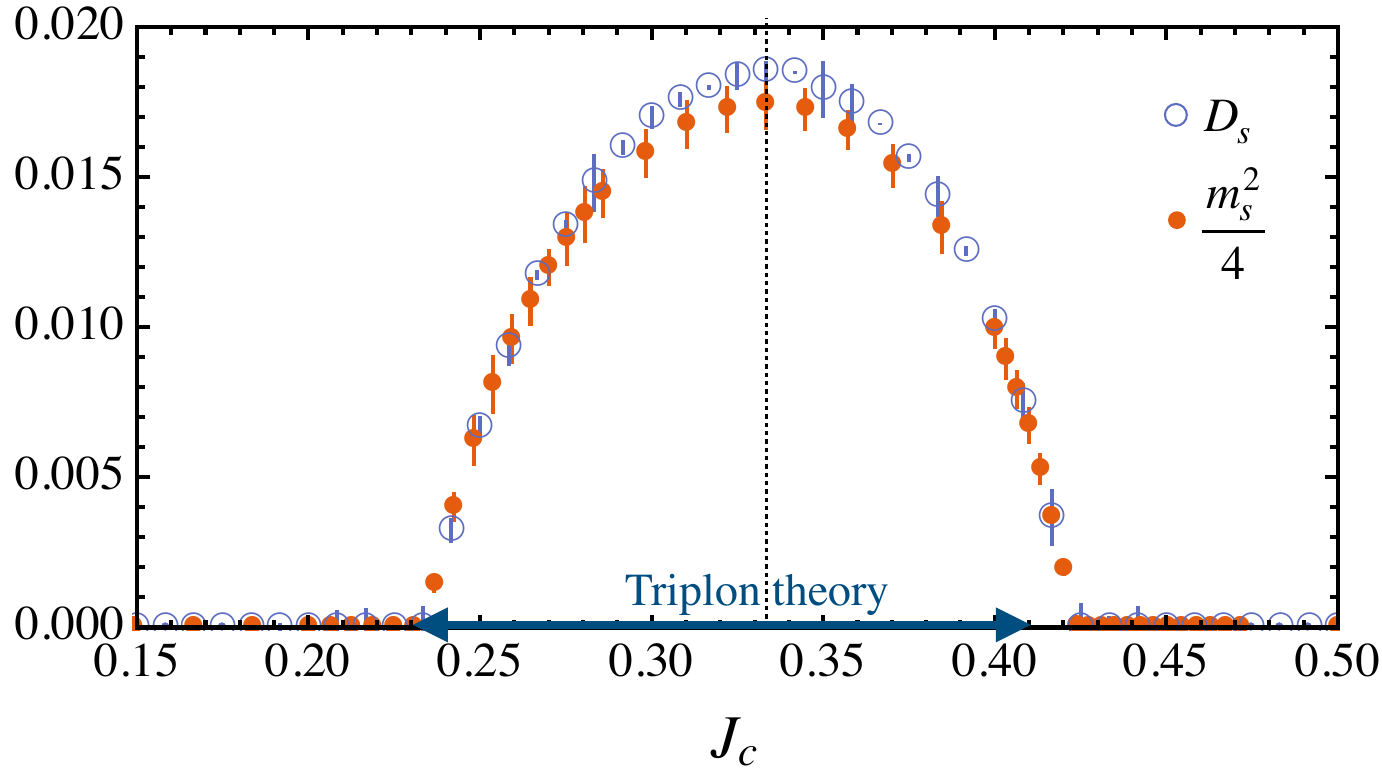}
\caption{Thermodynamic limit extrapolations of the spin stiffness, $D_s$, and staggered magnetization, $ m_s^2 $, from QMC simulations are plotted as a function of $J_C$ along the $J_A=J_B$ line. They produce a region of N\'eel phase that is consistent with triplon analysis (blue arrow).}
\label{fig:qmc-transition}
\end{figure}

In Fig.~\ref{fig:QPD-trip-stiff}, we present the stiffness data from our QMC calculations for a large lattice of $N = 864$ sites at a low temperature, $\beta = 1/T = 50$. Juxtaposed with the quantum phase diagram obtained from triplon analysis, the N\'eel phase obtained by spin stiffness exhibits remarkable agreement. The overall shape and extent of the region with $D_s \neq 0$ is not only qualitatively consistent with the phase boundary from triplon analysis, but it is also quantitative. This shows how good the proposed triplon description is for this model, even by such direct comparison with a large but finite size data.

We improve the phase boundary obtained from QMC by doing a systematic finite size scaling of $D_s$ and $m^2_s$ along the $J_A =J_B$ line.  Doing it for the whole phase diagram would be too tedious to extract their thermodynamic limit (TL) behaviours. While we consider an inverse temperature of $\beta=50$ for $D_s$, a slightly higher temperature of $\beta=20$ is taken for $\langle {\bf m}^2_s\rangle$ whose approach to TL is found to be slower (and harder) than that of $D_s$. The extrapolated values and error bars are obtained by the linear fits of $D_s$ and $m_s^2$ with respect to $1/ \sqrt{N}$~\cite{Sandvik-97}. These TL values of the two quantities, presented in Fig.~\ref{fig:qmc-transition} as a function of $J_C$, show an even closer agreement on the boundary of the N\'eel phase. When $J_C$ goes from 1/3 (centroid) to 0 (AB-hexagon side), the extrapolated values of both $D_s$ and $m_s$ go continuously to zero at { $J_C = 0.23(5)$}, which is precisely the critical point from the HS state triplon analysis. This is remarkable. The agreement is generally quite close along the black portion of the phase boundary in Fig.~\ref{fig:QPD-trip-stiff}. Across the red segments of the phase boundary (where the HS, DS and N\'eel phases all seem to be competing), the QMC estimate exceeds just a little beyond the phase boundary from theory. For example, along the $J_A=J_B$ line, as $J_C$ goes from 1/3 (centroid) to 1 (C-dimer corner), the extrapolated values of $D_s$ and $m^2_s$ vanish together at {$J_C=0.42(2)$}, only a little beyond the point $0.41$ on the red segment from theory. It is thus evident that the HS state triplon analysis provides a very good theory of this model to describe the thermodynamic properties, even if the tiny competing regions (not identified by our SSE calculations) leave room for some improvements.

\section{Magnetization Plateaus}\label{sec:Mag}
A notable feature of Cu$_2$(pymca)$_3$(ClO$_4$) is that its magnetization due to external magnetic field exhibits plateaus at $M/M_{sat}=1/3$ and 2/3~\cite{Okutani2019}. Of these, the plateau at 1/3 is  much wider compared to the one at 2/3. (We denote the magnetization along the field as $M$, and the saturated magnetization as $M_{sat}$.) Prompted by this behaviour,  we make a study of the magnetization in the ABC model on kagome-honeycomb lattice,
\begin{equation} 
\Hhat_{ABC} - h_{ext}\sum_{\vec{R}} \sum_{l=1}^6 S^z_{l,\vec{R}}\, , 
\label{eq:ABC-hext}
\end{equation}
 in the presence of an external magnetic field, $h_{ext}$. It correctly gives us the magnetization plateaus, reveals to us the underlying mechanism, and identifies the regions in the phase diagram in which either one or both plateaus occur; see Fig.~\ref{fig:mag-QPD}.

 \begin{figure}[t]
  	\centering
  	\includegraphics[width=0.46\textwidth]{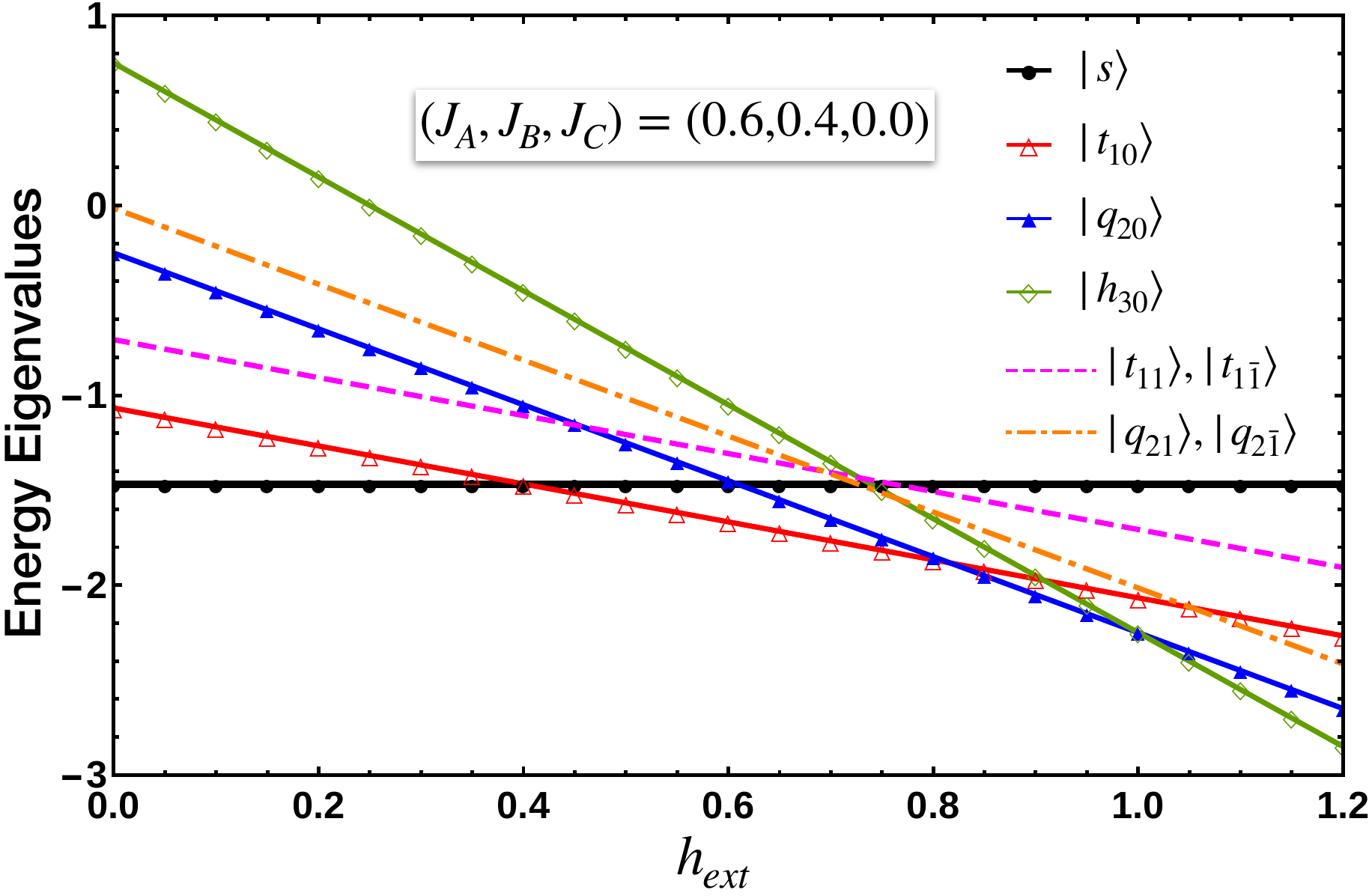}
  	\caption{Level-crossing diagram of an AB-hexagon. The energy levels with magnetic quantum number $m=1, 2,3$ cross the zero-field singlet ground state and amongst themselves with increasing magnetic field, $h_{ext}$. They form the basis for a theory of magnetization plateaus in Secs.~\ref{sec:1by3} and~\ref{sec:2by3}. The states with $m =-3,-2, -1, 0$, whose energies increase or stay constant with $h_{ext}$, are not quite relevant and not shown.}
  	\label{fig:eig-AB-h}
  \end{figure}

Consider first the eigenstates of a single AB-hexagon.  Figure~\ref{fig:eig-AB-h} shows how they compete as a function of $h_{ext}$. The most notable feature of this level-crossing diagram is that, as $h_{ext}$ is increased from zero, the ground state of the hexagon successively changes from a singlet, $|s\rangle$, to a triplet, $|t_{10}\rangle$, to a quintet, $|q_{20}\rangle$, to the fully polarized heptet state, $|h_{30}\rangle$; see Appendix~\ref{app:eigHex} for the hexagon's eigenstates. Correspondingly, the $M/M_{sat}$ of the hexagon increases in steps from 0 to 1/3 to 2/3 to 1. Notice that $|t_{10}\rangle$ remains the ground state over a wider range of $h_{ext}$ as compared to $|q_{20}\rangle$, i.e., the magnetization stays at 1/3 over a wider range of the magnetic field as compared to 2/3. All this is remarkably like the plateaus observed in Cu$_2$(pymca)$_3$(ClO$_4$). Hence, for Eq.~\eqref{eq:ABC-hext}, we derive and study the effective models in terms of the hexagonal eigenstates relevant for $M/M_{sat}=$ 0, 1/3, 2/3, 1. 

For the non-magnetic case,  the reference state is the hexagonal singlet state, $\ket{s}$. It was the reference state for triplon analysis in Sec.~\ref{sec:HS}. We do the same analysis again, but with a non-zero $h_{ext}$. It gives us the critical field at which the non-magnetic ground state gives way to non-zero magnetization.  
The fully saturated magnetic state is a trivial eigenstate of the ABC model. Its stability against a spin-flip excitation determines the saturation field, which turns out to be $J_A + J_B +J_C$. The theories of the magnetization plateaus at 1/3 and 2/3 are presented in the following subsections. 

\subsection{Theory of 1/3 plateau}\label{sec:1by3}
In the independent hexagon limit of the ABC model, say for $J_C=0$, the 1/3 plateau is described by a unique state  wherein every AB hexagon is in the triplet state, $|t_{10}\rangle$. For a non-zero $J_C$, this ideal reference state would quantum fluctuate and get renormalized. Thus, the 1/3 plateau would exist as long as the energy gap to these fluctuations is non-zero. The minimal set of hexgonal eigenstates required to do a theory of the 1/3 plateau is $\{|s\rangle, |t_{10}\rangle, |q_{20}\rangle \}$. It can be enlarged by also including $\ket{q_{21}}$ and $\ket{q_{2\bar{1 }}}$ from the quintets in $\nu=1$, $\bar{1}$ sectors (see Appendix~\ref{app:eigHex}), which interact directly with $\ket{t_{10}}$. It improves the result slightly, specially around $J_A=J_B$ line; qualitatively, the two give same results. 

As in Sec.~\ref{sec:trip}, we associate  boson operators $\shat^\dag_{\vec{R}}$, $\that^\dag_{10, \vec{R}}$ and $\qhat^\dag_{2\nu,\vec{R}}$ to the respective kets of the AB-hexagon at position $\vec{R}$. With a simplifying approximation, $\that_{10,\vec{R}} \approx \tbar$, we describe the reference state by a mean amplitude, $\tbar$, for every AB-hexagon to be in the state $\ket{t_{10}}$. Thus, the AB part of the model in Eq.~\eqref{eq:ABC-hext}, including the magnetic field term, can be written as: $ \Hhat_{A} + \Hhat_{B} - h_{ext}\sum_{\vec{R}} \sum_{l=1}^6 S^z_{l,\vec{R}} \approx \sum_{\vec{R}}\Big[E_s \hat{s}_{\vec{R}}^\dag \hat{s}_{\vec{R}} + (E_{t0} - h_{ext})\bar{t}^2 + (E_{q0} -2h_{ext})\hat{q}_{20,\vec{R}}^\dag \hat{q}_{20,\vec{R}} + (E_{q1} -2h_{ext})(\hat{q}_{21,\vec{R}}^\dag \hat{q}_{21,\vec{R}} + \hat{q}_{2\bar{1},\vec{R}}^\dag \hat{q}_{2\bar{1},\vec{R}})\Big]$, where $E_s$, $E_{t0}$, $E_{q\nu}$ (for $\nu=0,1,\bar{1}$) denote the eigenvalues of $\ket{s}$, $\ket{t_{10}}$, $\ket{q_{2\nu}}$ respectively, and $E_{q\bar{1}}=E_{q1}$. We also add to it $\lambda\sum_{\vec{R}}( \bar{t}^2+\hat{s}_{\vec{R}}^\dag \hat{s}^{ }_{\vec{R}} + \sum_{\nu}\hat{q}_{2\nu,\vec{R}}^\dag\hat{q}^{ }_{2\nu,\vec{R}}-1)$ to meet the constraint on average through Lagrange multiplier $\lambda$. The interaction between the AB-hexagons through $\Hhat_C$ is expressed using the representation in Eq.~\eqref{eq:rep1by3} for the spins of every AB-hexagon. Putting these together in Eq.~\eqref{eq:ABC-hext}, and doing the Fourier transformation: $\hat{s}_{\vec{R}}=\frac{1}{\sqrt{N_{uc}}}\sum_{\vec{k}}\hat{s}_{\vec{k}} \, e^{i \vec{k}\cdot \vec{R}}$ and $\hat{q}_{2\nu,\vec{R}}=\frac{1}{\sqrt{N_{uc}}} \sum_{\vec{k}}\hat{q}_{2\nu,\vec{k}} \, e^{i \vec{k}\cdot \vec{R}}$, we get the following effective Hamiltonian for 1/3 plateau:
\begin{equation}
	\Hhat^{(\frac{1}{3})}= \epsilon_0^{(\frac{1}{3})}N_{uc} + \sum_{\vec{k}}\Psi_{\vec{k}}^{(\frac{1}{3})\dag} \, \mathcal{H}_{\vec{k}}^{(\frac{1}{3})} \, \Psi_{\vec{k}}^{(\frac{1}{3})}.
	\label{eq:H1by3}
\end{equation}	
For $\epsilon_0^{(\frac{1}{3})}$ and $\mathcal{H}_{\vec{k}}^{(\frac{1}{3})}$, see Appendix~\ref{app:1by3}. The $\Psi_{\vec{k}}^{(\frac{1}{3})}$ is a Nambu column vector whose adjoint, $\Psi_{\vec{k}}^{(\frac{1}{3})\dag}$, is given below.
\begin{equation}
	\Psi_{\vec{k}}^{(\frac{1}{3})\dag} = \begin{pmatrix}
		\hat{s}^\dag_{\vec{k}} & \hat{q}^\dag_{20,\vec{k}} & \hat{q}^\dag_{21,\vec{k}} & \hat{q}^\dag_{2\bar{1},\vec{k}} & \hat{s}_{\vec{-k}} & \hat{q}_{20,\vec{-k}} & \hat{q}_{21,\vec{-k}} & \hat{q}_{2\bar{1},\vec{-k}}
	\end{pmatrix}
\end{equation}

The Bogoliubov diagonalization of Eq.~\eqref{eq:H1by3} gives four quasiparticle dispersions, $2\epsilon_{j,\vec{k}}^{(\frac{1}{3})}$. The ground state energy per unit-cell of $\Hhat^{(\frac{1}{3})}$ can be written as: $\mathcal{E}_{g}^{(\frac{1}{3})} = \epsilon_0^{(\frac{1}{3})} + \frac{1}{N_{uc}}\sum_{\vec{k}} \sum_{j=1}^4 \epsilon_{j,\vec{k}}^{(\frac{1}{3})}$. Minimizing $\mathcal{E}_g^{(\frac{1}{3})}$ with respect to $\bar{t}^2$ and $\lambda$ gives the following self-consistent equations.
\begin{subequations}
\label{eq:sc1by3}
\begin{align}
\lambda &= h_{ext}  -E_{t0} + \frac{3}{2} J_C \chi  -\frac{1}{N_{uc}}\sum_{\vec{k}}\sum_{j=1}^4\frac{\partial \epsilon_{j,\vec{k}}^{(\frac{1}{3})}}{\partial \bar{t}^2}\\
\bar{t}^2 &= 3 -\frac{1}{N_{uc}}\sum_{\vec{k}}\sum_{j=1}^4\frac{\partial \epsilon_{j,\vec{k}}^{(\frac{1}{3})}}{\partial \lambda}
\end{align}
\end{subequations}	
By solving these equations for $\lambda$ and $\tbar^2$, we determine the quasiparticle energy gap, and hence the 1/3 plateau. The results from this calculation are discussed in Sec.~\ref{sec:mag-qpd}.
  
\subsection{Theory of 2/3 plateau}\label{sec:2by3}
We can do a minimal theory of 2/3 plateau in terms of the states $\{\ket{t_{10}}, \ket{q_{20}}, \ket{h_{30}}\}$, or a more general one by also considering two other triplet states, $\ket{t_{11}}$ and $\ket{t_{1\bar{1}}}$. The plateau region obtained from both the calculations is pretty much the same. So, we describe only 
the minimal theory. Let $\hat{t}^\dag_{10}, \hat{q}^\dag_{20}$, $\hat{h}^\dag_{30}$ be the boson operators corresponding to the kets $\ket{t_{10}}, \ket{q_{20}}$, $\ket{h_{30}}$ respectively. With $\ket{q_{20}}$ as the reference state on 2/3 plateau, we approximate $\hat{q}_{20}$ by a mean amplitude $\bar{q}$. Thus, in Eq.~\eqref{eq:ABC-hext}, $\Hhat_A + \Hhat_B - h_{ext}\sum_{\vec{R},l}S^z_{l,\vec{R}} + {\rm constraint} \approx [(E_{q0}-2 h_{ext} +\lambda) \bar{q}^2 - \lambda] N_{uc} + \sum_{\vec{R}} [(E_{t0}-h_{ext}+\lambda) \that^\dag_{10,\vec{R}} \that^{ }_{10,\vec{R}} + (E_{h0}-3h_{ext}+\lambda) \hhat^\dag_{30,\vec{R}} \hhat^{ }_{30,\vec{R}}]$; $\lambda$ is the Lagrange multiplier. We write $\Hhat_C$ using  Eq.~\eqref{eq:rep2by3mini}. The final effective Hamiltonian in the $\vec{k}$-space describing triplon and hepton fluctuations with respect to the 2/3 plateau can be written as: 
\begin{align}
		\Hhat^{(\frac{2}{3})} =& ~ \sum_{\vec{k}}\Big[D_{t, \vec{k}} \, \hat{t}_{10,\vec{k}}^\dag \hat{t}_{10,\vec{k}} + D_{h,\vec{k}} \, \hat{h}_{h,\vec{k}} \hat{h}^\dag_{30,\vec{k}} \nonumber \\
		& ~ + F_{\vec{k}}\big(\hat{t}_{10,\vec{k}}^\dag \hat{h}_{30,-\vec{k}}^\dag + h.c.\big)\Big] + \epsilon_0^{(\frac{2}{3})}  N_{uc}
\end{align}
where $D_{t,\vec{k}}$, $D_{h,\vec{k}}$, $F_{\vec{k}}$ and $\epsilon_0^{(\frac{2}{3})}$ are given in Eqs.~\eqref{eq:DDFeps}.	

Diagonalization of $\Hhat^{(\frac{2}{3})}$ gives the following two quasiparticle dispersions:
\begin{equation}
\epsilon^{(\pm)}_{\vec{k}} = \pm \frac{(D_{h,\vec{k}}-D_{t,\vec{k}})}{2} + \sqrt{\frac{(D_{t,\vec{k}} + D_{h,\vec{k}})^2}{4} - F_{\vec{k}}^2} \,. 
\label{eq:dis2by3}
\end{equation}	
The ground state energy per unit-cell of $\Hhat^{(\frac{2}{3})}$ is given by $ \mathcal{E}^{(\frac{2}{3})}_g = \epsilon_0^{(\frac{2}{3})} + \frac{1}{N_{uc}}\sum_{\vec{k}}\epsilon^{(+)}_{\vec{k}}$. Its minimization with respect to $\lambda$ and $\bar{q}^2$ leads to the equations
\begin{subequations}
\label{eq:sc2by3}
\begin{align}
\lambda &= 2h_{ext} -E_{q0} -3 J_C \tilde{\chi} -\frac{1}{N_{uc}}\sum_{\vec{k}}\frac{\partial \epsilon^{(+)}_{\vec{k}}}{\partial \bar{q}^2}\\
\bar{q}^2 &= 2-\frac{1}{N_{uc}}\sum_{\vec{k}}\frac{\partial \epsilon^{(+)}_{\vec{k}}}{\partial \lambda}
\end{align}
\end{subequations}
whose self-consistent solution gives the region of 2/3 magnetization plateau described below. 

\subsection{Results and implications for Cu$_2$(pymca)$_3$(ClO$_4$)}\label{sec:mag-qpd}
For a given $(J_A,J_B,J_C)$ in the ternary phase diagram (refer to Fig.~\ref{fig:QPD-trip}), we compute the quasiparticle energy gap for the 1/3 plateau by solving Eqs.~\eqref{eq:sc1by3} for different values of $h_{ext}$, and find the range of $h_{ext}$ over which this energy gap stays non-zero. As long as this range has a finite width, we have a 1/3 plateau. But when it shrinks to zero, the 1/3 plateau ceases to exist. By scanning over the ternary diagram and the magnetic field, we obtain the region of existence of the 1/3 plateau. We do likewise for the 2/3 plateau by solving Eqs.~\eqref{eq:sc2by3}. 

\begin{figure}[t]
  	\centering
  	\includegraphics[width=0.48\textwidth]{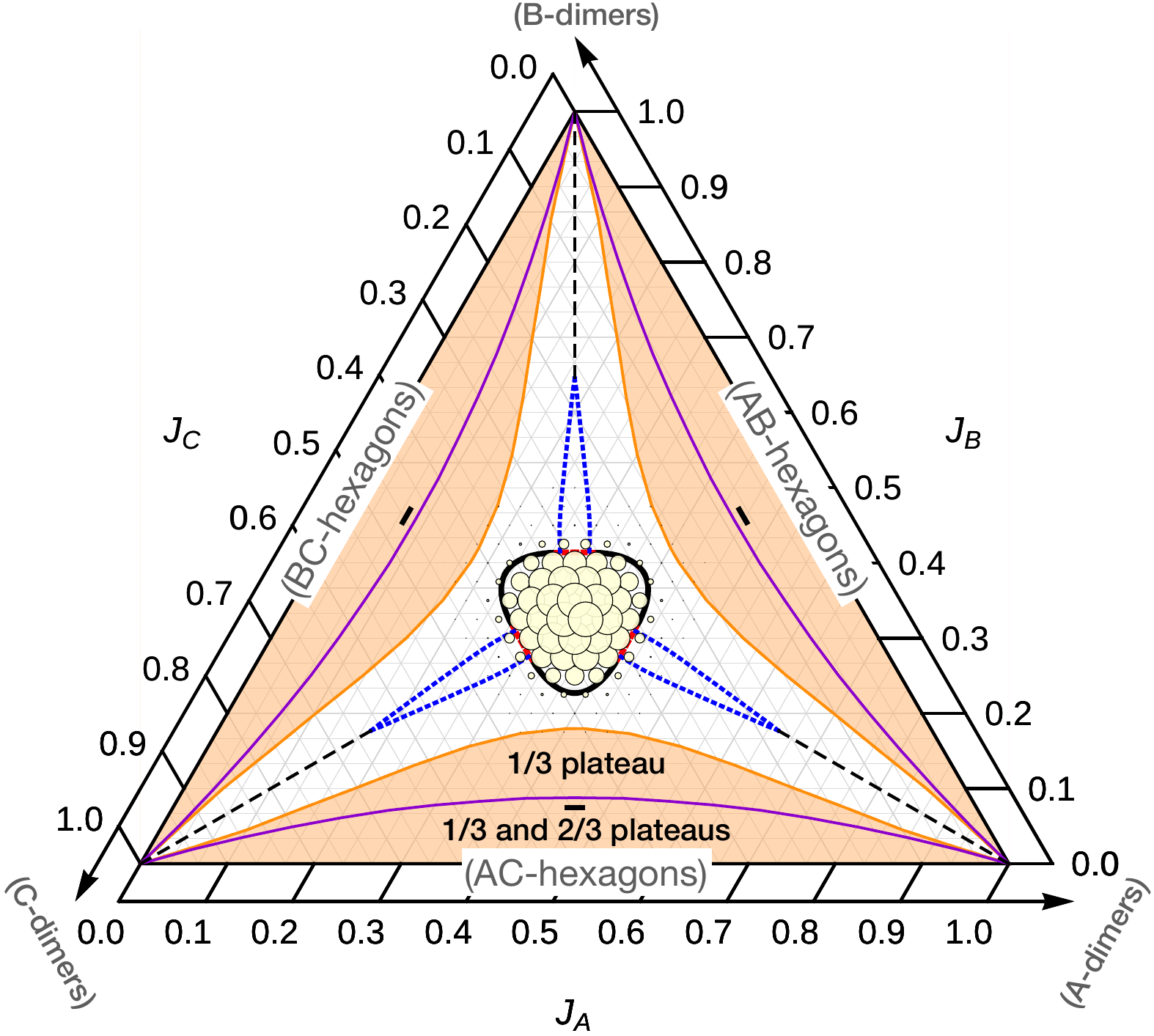}
  	\caption{Regions of existence of the magnetization plateaus in the phase diagram. 
	The 1/3 plateau occurs everywhere inside the regions filled with orange colour. No magnetization plateaus occur outside these regions. Inside the 1/3 plateau regions, below the purple lines, the 2/3 plateau also exists. The little black marks just below the purple lines denote the estimated position of Cu$_2$(pymca)$_3$(ClO$_4$).}
  	\label{fig:mag-QPD}
\end{figure}

The regions of existence of the magnetization plateaus thus obtained are shown in Fig.~\ref{fig:mag-QPD}. The 1/3 plateau is found to exist inside the orange-coloured bounded regions adjoining the three sides of the ternary diagram. For instance, along $J_A=J_B$ line, the 1/3 plateau exists for $0 \le J_C \lesssim 0.18$; along $J_B=0.2$ line, it occurs for $0 \le J_C \lesssim 0.102$ and $0.698 \lesssim J_C \le 0.8$. Inside these regions of the 1/3 plateau, we also find the 2/3 plateau to occur in the smaller regions adjacent to the sides of the ternary diagram, bounded by the arc-shaped purple lines, as shown in Fig.~\ref{fig:mag-QPD}. Along $J_A=J_B$ line, the 2/3 plateau occurs for $J_C$ between 0 and 0.088; along $J_B=0.2$, it occurs for $0 \le J_C \lesssim 0.061$ and $0.739 \lesssim  J_C  \le 0.8$.

A notable feature of our findings is that the 2/3 plateau always occurs with 1/3 plateau, or the 1/3 plateau alone exists. It puts a constraint on the exchange interactions in Cu$_2$(pymca)$_3$(ClO$_4$), which exhibits both the plateaus. We also find the width of the 2/3 plateau to be always smaller than that of the 1/3 plateau, consistent with the observed behaviour in Cu$_2$(pymca)$_3$(ClO$_4$). Note that the region of 1/3 plateau lies strictly inside the zero-field hexagonal-singlet phase. This is an interesting fact of our theory, which unambiguously implies that, because Cu$_2$(pymca)$_3$(ClO$_4$) exhibits 1/3 magnetization plateau, therefore in the absence of magnetic field, it must have the hexagonal-singlet ground state. 

\begin{figure}[htbp]
  	\centering
  	\includegraphics[width=0.48\textwidth]{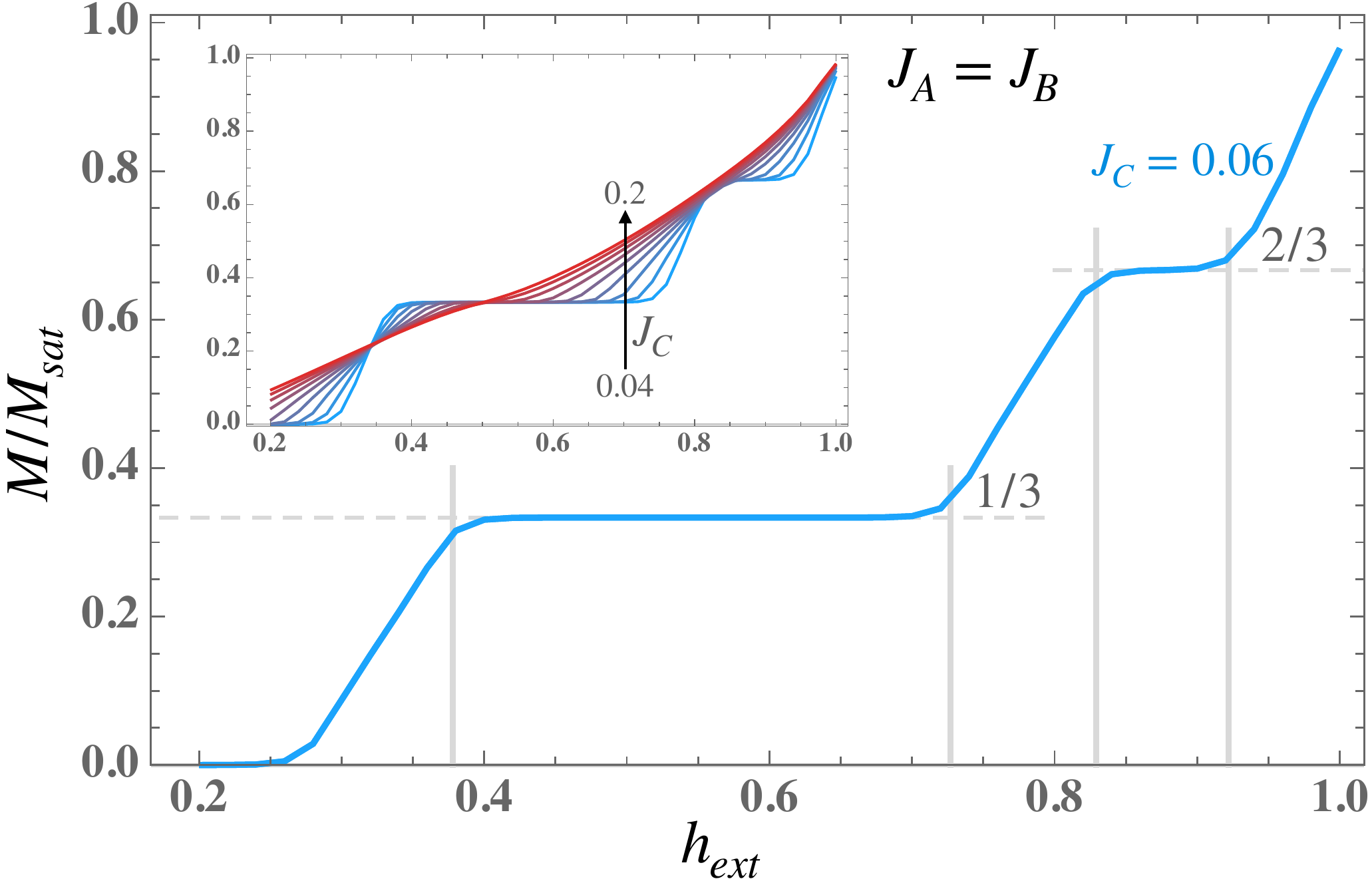}
  	\caption{Magnetization versus magnetic field along $J_A=J_B$ line in the phase diagram. The blue curve with plateaus at 1/3 and 2/3 is the QMC data for $J_C=0.06$, and the vertical grey lines are the critical fields from theory. (Inset) Evolution of the plateaus with $J_C$ increasing from 0.04 to 0.2.}
  	\label{fig:mag-qmc}
  \end{figure}

We check these findings by doing QMC simulations of the ABC model in magnetic field. In Fig.~\ref{fig:mag-qmc}, we present the QMC data for 384 spins at a low enough temperature ($\beta=100$) along $J_A=J_B$ line. The inset of this figure shows the evolution of $M$ vs. $h_{ext}$ with $J_C$. For smaller $J_C$ values, our QMC data exhibits plateaus at 1/3 as well as 2/3. Upon increasing $J_C$, first the 2/3 plateau tends to vanish around 0.09 and then the 1/3 plateau disappears around 0.185, in agreement with our theory. The positions and the widths of the plateaus obtained from QMC simulations are also compared with the critical fields calculated from theory. One such comparison for $J_C=0.06$ presented in Fig.~\ref{fig:mag-qmc} looks pretty good. We have made similar checks also along directions other than $J_A=J_B$, and the QMC numerics is found to be consistent with the theory.

\begin{figure}[t]
\includegraphics[width=0.46\textwidth]{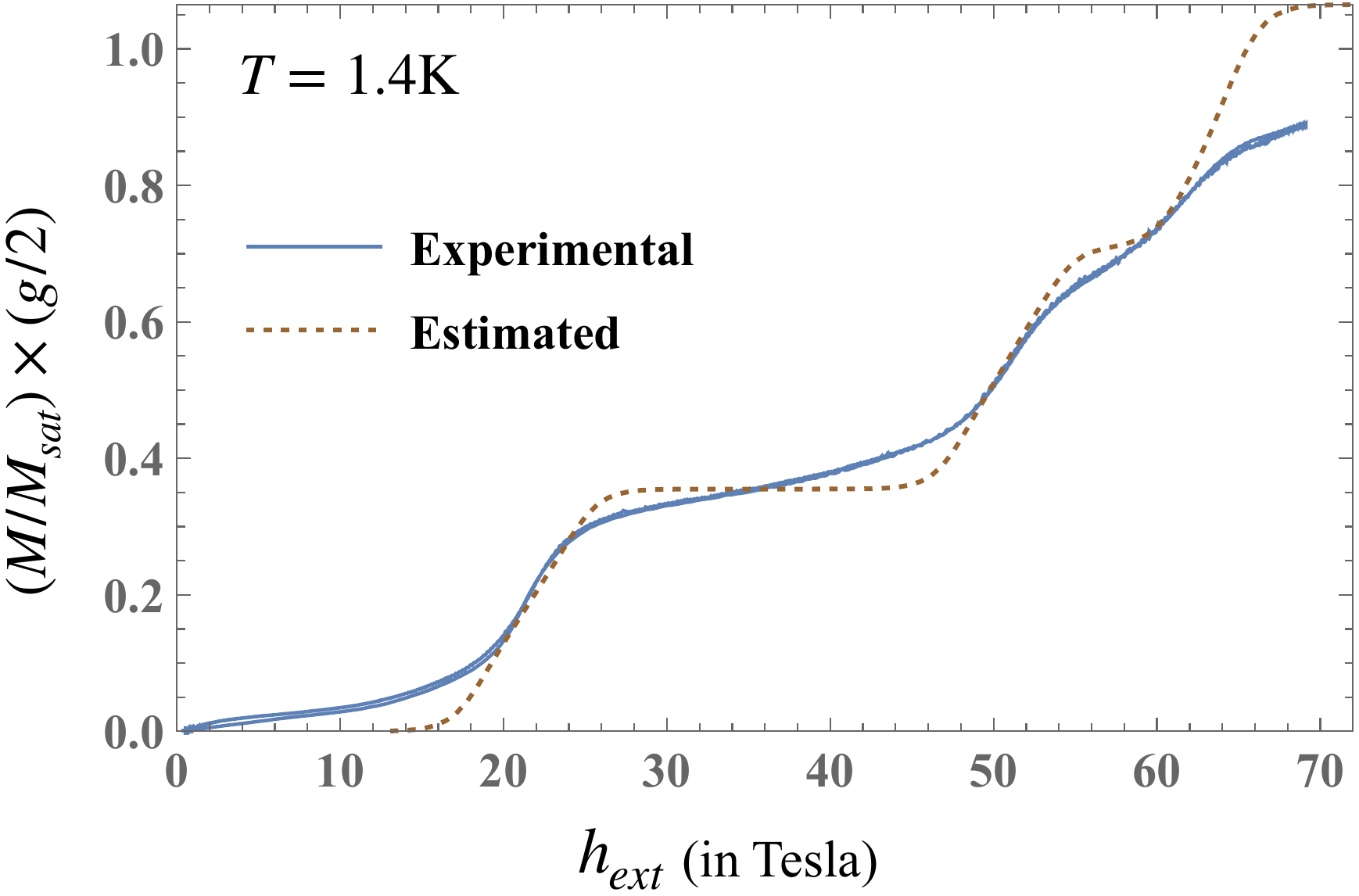}
\caption{Comparison of the magnetization measured for Cu$_2$(pymca)$_3$(ClO$_4$) (the experimental data of Ref.~\cite{Okutani2019}) with the magnetization calculated  by us using QMC method at 1.4K for the estimated interactions, $(J_A,J_B,J_C)=(43.7\pm0.8, 43.7 \mp 0.8, 7.1)$K. The calculated magnetization is multiplied by $g/2$ (with a Land\'e $g$-factor of $g=2.13$ for the material) to have the same $M_{sat}$ as for the measured data.}
\label{fig:theoexp}
\end{figure}

We also estimate the exchange interactions for Cu$_2$(pymca)$_3$(ClO$_4$), and find that  the experimental data of magnetization in Ref.~\cite{Okutani2019} is best described for $J_C = 0.075$ and $J_A \approx J_B$, with an interaction strength of $J_A+J_B+J_C \approx 66$T (94.4K). See Fig.~\ref{fig:theoexp} for a comparison of the experimental data with the magnetization produced by our QMC simulations for the estimated interactions of values $J_C= 5$T (7.1K) and $J_A=J_B = 30.5$T (43.7K); the two compare nicely with a particularly good match along the ramps on either side of the 1/3 plateau. This value of $J_A (J_B)$ is same as considered in Ref.~\cite{Okutani2019}, but our slightly weaker value of $J_C$ presents a better match~\footnote{Ref.~\cite{Okutani2019} overestimates $J_C$. In our dimensionless convention, their $J_C$ is equal to $0.091$, which for $J_A=J_B$ barely lies on the boundary of the 2/3 plateau region.}.
Note that the other closeby estimates such as $(J_A,J_B,J_C) = (0.471, 0.454, 0.075) \equiv (31, 30, 5)$T$\equiv (44.5, 42.9, 7.1)$K also produce the same match, but going farther away from the $J_A=J_B$ line clearly spoils it. The choices of $J_C=0.075$ and $J_A+J_B+J_C=66$T are found to be less flexible in search for the best match, and so are our best choices.

Notably, this estimate puts Cu$_2$(pymca)$_3$(ClO$_4$) just inside the region of two plateaus, close to its boundary with the one plateau region; the black marks just below the purple lines in Fig.~\ref{fig:mag-QPD} denote the estimated position(s) of this material in the phase diagram. It makes the 2/3 plateau in Cu$_2$(pymca)$_3$(ClO$_4$) highly susceptible to small changes in the interactions, and points to a real possibility of making the 2/3 plateau disappear continuously, say, by applying pressure. This is an interesting prediction for the experimentalists to investigate.

While the key features of the magnetization behaviour of Cu$_2$(pymca)$_3$(ClO$_4$) are described well by the antiferromagnetic Heisenberg model on kagome-honeycomb lattice, the following differences visible in Fig.~\ref{fig:theoexp} still remain to be understood. The experimental data does not saturate even upto a field of 70T where the calculated magnetization at 1.4K saturates. The experimental magnetization exhibits a slow but steady growth well before 17T (the estimated critical field where the non-magnetic state gives way to magnetization; it corresponds to the zero-field spin-gap of 24.5K). This conspicuous variation of magnetization is also seen on the 1/3 plateau; presumably the same also weakens the already small 2/3 plateau. Moreover, a plateau-like tendency is noted above 2/3. These differences between the measured and the calculated magnetization suggest that there are other interactions at work in this compound, in addition to but subdominant to the exchange interactions considered here.

\section{Conclusion}\label{sec:sum}
The quantum phase diagram of an antiferromagnetic spin-1/2 Heisenberg model on kagome-honeycomb lattice is obtained by a combined study based on triplon analysis and QMC simulations. The findings from the two approaches are mutually consistent both qualitatively and quantitatively. Interestingly, while the model is unfrustrated and bipartite, its phase diagram is dominated by a quantum paramagnetic phase that is best described as hexagonal singlet state. The N\'eel antiferromagnetic order appears only in a small region around the uniform honeycomb case. 
The model is studied further in an external magnetic field to understand the magnetization behaviour observed in Cu$_2$(pymca)$_3$(ClO$_4$). To this end, a theory of the magnetization plateaus is developed and confirmed by the QMC simulations. It leads to identifying the regions of one (1/3), two (1/3 and 2/3) or no plateaus in the phase diagram, and discovers an existential relation between the plateaus and the zero-field hexagonal singlet ground state. The occurrence of 1/3 plateau in Cu$_2$(pymca)$_3$(ClO$_4$) is thus a proof that this compound has a gapped hexagonal-singlet ground state in the absence of the magnetic field. An estimation of the exchange interactions places Cu$_2$(pymca)$_3$(ClO$_4$) near the boundary of the two-plateau phase. It implies that a small application of pressure (or another non-thermal variable that may effect some change to the exchange interactions) may cause the disappearance of the 2/3 plateau. Thus, Cu$_2$(pymca)$_3$(ClO$_4$) presents a scope for investigating a quantum phase transition from the two-plateaus to one-plateau phase.

\begin{acknowledgments}
M.A. acknowledges DST (India) for INSPIRE fellowship, and thanks Pratyay Ghosh for discussions. B.K. acknowledges SERB (India) research grant for project No. CRG/2019/003251. We also acknowledge the DST-FIST-funded HPC facility at the School of Physical Sciences, JNU for computations. We thank Masayuki Hagiwara for sharing their magnetization data of Ref.~\cite{Okutani2019}.
\end{acknowledgments}

\appendix

\section{Heisenberg problem and triplon representation on a single AB-hexagon}{\label{app:eigHex}}
The Hamiltonian of a single spin-1/2 AB hexagon can be written as:
\begin{align}
\hat{h}_{AB} =&~ J_A\left(\vec{S}_{2}\cdot\vec{S}_{3} + \vec{S}_{4}\cdot\vec{S}_{5}+\vec{S}_{6}\cdot\vec{S}_{1}\right) \nonumber \\
		& +J_B\left(\vec{S}_{1}\cdot\vec{S}_{2}+\vec{S}_{3}\cdot\vec{S}_{4}+\vec{S}_{5}\cdot\vec{S}_{6}\right).  
	\label{eq:ABhamil}
\end{align}
The total spin, $S_{total}$, and its $z$-component, $S^z_{total}$, are two conserved quantities of this Hamiltonian. Let the quantum number corresponding to $S^z_{total}$ be $m$, in terms of which the Hilbert space of six spin-1/2's, $\{|\uparrow\rangle,|\downarrow\rangle\}^{\otimes 6}$, can be sectorized into seven parts for $m=0,\pm1,\pm2,\pm3$. Here, $\ket{\uparrow}$ and $\ket{\downarrow}$ are the eigenstates of an individual $S^z$ operator, with eigenvalues $\frac{1}{2}$ and $-\frac{1}{2}$ respectively. This Hamiltonian also has a threefold rotational symmetry, $\mathcal{R}_{\frac{2\pi}{3}}$. Furthermore, $\comm{S^z_{total}}{\mathcal{R}_{\frac{2\pi}{3}}}=0$. Hence, the basis states in each fixed $m$ sector can be further grouped into smaller sectors using the rotational quantum number, $\nu=0,1,\bar{1} $ corresponding respectively to the threefold rotation eigenvalues $1,\omega,\omega^2$.  Here, $\bar{1}$ stands for $-1$. We write the $\hat{h}_{AB}$ in matrix form in each of these $(m,\nu)$ subspaces separately, and find the complete eigenspectrum for different values of $J_B/J_A$ varying from 0 to 1. 

\begin{table}[t]
	\caption{}
	\label{tab:m0}
	\centering 
	\begin{tabular}{c | c} 
		\hline
		 $\nu$ & Basis states for $m=0$ \\ [1ex] \hline 
		&~ $\ket{\uparrow\downarrow\uparrow\downarrow\uparrow\downarrow}$,~$\ket{\downarrow\uparrow\downarrow\uparrow\downarrow\uparrow}$,\\
		0 &~$\frac{1}{\sqrt{3}}(\ket{\uparrow\uparrow\uparrow\downarrow\downarrow\downarrow}+\ket{\uparrow\downarrow\downarrow\downarrow\uparrow\uparrow}+\ket{\downarrow\downarrow\uparrow\uparrow\uparrow\downarrow})$,\\[0.8ex]
		&~$\frac{1}{\sqrt{3}}(\ket{\uparrow\uparrow\downarrow\uparrow\downarrow\downarrow}+\ket{\downarrow\uparrow\downarrow\downarrow\uparrow\uparrow}+\ket{\downarrow\downarrow\uparrow\uparrow\downarrow\uparrow})$,\\[0.8ex]
		&~$\frac{1}{\sqrt{3}}(\ket{\downarrow\downarrow\uparrow\downarrow\uparrow\uparrow}+\ket{\uparrow\downarrow\uparrow\uparrow\downarrow\downarrow}+\ket{\uparrow\uparrow\downarrow\downarrow\uparrow\downarrow})$,\\[0.8ex]
		&$\frac{1}{\sqrt{3}}(\ket{\uparrow\downarrow\downarrow\uparrow\downarrow\uparrow}+\ket{\downarrow\uparrow\downarrow\uparrow\uparrow\downarrow}+\ket{\downarrow\uparrow\uparrow\downarrow\downarrow\uparrow})$, \\[0.8ex]
		&~$\frac{1}{\sqrt{3}}(\ket{\downarrow\downarrow\downarrow\uparrow\uparrow\uparrow}+\ket{\downarrow\uparrow\uparrow\uparrow\downarrow\downarrow}+\ket{\uparrow\uparrow\downarrow\downarrow\downarrow\uparrow})$,\\[0.8ex]
		&~$\frac{1}{\sqrt{3}}(\ket{\downarrow\uparrow\uparrow\downarrow\uparrow\downarrow}+\ket{\uparrow\downarrow\uparrow\downarrow\downarrow\uparrow}+\ket{\uparrow\downarrow\downarrow\uparrow\uparrow\downarrow})$ \\[1ex] \hline
		
		&~$\frac{1}{\sqrt{3}}(\omega\ket{\uparrow\uparrow\uparrow\downarrow\downarrow\downarrow}+\omega^2\ket{\uparrow\downarrow\downarrow\downarrow\uparrow\uparrow}+\ket{\downarrow\downarrow\uparrow\uparrow\uparrow\downarrow})$,\\[0.8ex]
		1&$\frac{1}{\sqrt{3}}(\omega\ket{\uparrow\uparrow\downarrow\uparrow\downarrow\downarrow}+\omega^2\ket{\downarrow\uparrow\downarrow\downarrow\uparrow\uparrow}+\ket{\downarrow\downarrow\uparrow\uparrow\downarrow\uparrow})$,\\[0.8ex]
		&~$\frac{1}{\sqrt{3}}(\omega\ket{\downarrow\downarrow\uparrow\downarrow\uparrow\uparrow}+\omega^2\ket{\uparrow\downarrow\uparrow\uparrow\downarrow\downarrow}+\ket{\uparrow\uparrow\downarrow\downarrow\uparrow\downarrow})$,\\[0.8ex]
		&$\frac{1}{\sqrt{3}}(\omega\ket{\uparrow\downarrow\downarrow\uparrow\downarrow\uparrow}+\omega^2\ket{\downarrow\uparrow\downarrow\uparrow\uparrow\downarrow}+\ket{\downarrow\uparrow\uparrow\downarrow\downarrow\uparrow})$, \\[0.8ex]
		&~$\frac{1}{\sqrt{3}}(\omega\ket{\downarrow\downarrow\downarrow\uparrow\uparrow\uparrow}+\omega^2\ket{\downarrow\uparrow\uparrow\uparrow\downarrow\downarrow}+\ket{\uparrow\uparrow\downarrow\downarrow\downarrow\uparrow})$,\\[0.8ex]
		&~$\frac{1}{\sqrt{3}}(\omega\ket{\downarrow\uparrow\uparrow\downarrow\uparrow\downarrow}+\omega^2\ket{\uparrow\downarrow\uparrow\downarrow\downarrow\uparrow}+\ket{\uparrow\downarrow\downarrow\uparrow\uparrow\downarrow})$ \\[1ex] \hline

		&~$\frac{1}{\sqrt{3}}(\omega^2\ket{\uparrow\uparrow\uparrow\downarrow\downarrow\downarrow}+\omega\ket{\uparrow\downarrow\downarrow\downarrow\uparrow\uparrow}+\ket{\downarrow\downarrow\uparrow\uparrow\uparrow\downarrow})$,\\[0.8ex]
		$\bar{1}$&$\frac{1}{\sqrt{3}}(\omega^2\ket{\uparrow\uparrow\downarrow\uparrow\downarrow\downarrow}+\omega\ket{\downarrow\uparrow\downarrow\downarrow\uparrow\uparrow}+\ket{\downarrow\downarrow\uparrow\uparrow\downarrow\uparrow})$,\\[0.8ex]
		&~$\frac{1}{\sqrt{3}}(\omega^2\ket{\downarrow\downarrow\uparrow\downarrow\uparrow\uparrow}+\omega\ket{\uparrow\downarrow\uparrow\uparrow\downarrow\downarrow}+\ket{\uparrow\uparrow\downarrow\downarrow\uparrow\downarrow})$,\\[0.8ex]
		&~$\frac{1}{\sqrt{3}}(\omega^2\ket{\uparrow\downarrow\downarrow\uparrow\downarrow\uparrow}+\omega\ket{\downarrow\uparrow\downarrow\uparrow\uparrow\downarrow}+\ket{\downarrow\uparrow\uparrow\downarrow\downarrow\uparrow})$, \\[0.8ex]
		&~$\frac{1}{\sqrt{3}}(\omega^2\ket{\downarrow\downarrow\downarrow\uparrow\uparrow\uparrow}+\omega\ket{\downarrow\uparrow\uparrow\uparrow\downarrow\downarrow}+\ket{\uparrow\uparrow\downarrow\downarrow\downarrow\uparrow})$,\\[0.8ex]
		&~$\frac{1}{\sqrt{3}}(\omega^2\ket{\downarrow\uparrow\uparrow\downarrow\uparrow\downarrow}+\omega\ket{\uparrow\downarrow\uparrow\downarrow\downarrow\uparrow}+\ket{\uparrow\downarrow\downarrow\uparrow\uparrow\downarrow})$ \\[1ex]
		\hline 
	\end{tabular}
\end{table}

The ground state of $\hat{h}_{AB}$ is a nondegenerate unique singlet state (i.e., $S_{tot}=0$) in the 8-dimensional $(m,\nu)=(0,0)$ subspace. See Table~\ref{tab:m0} for the basis states with $m=0$. Let us denote this state as $\ket{s}$ and call the corresponding ground state energy as $E_{s}$. 

The first excited state of $\hat{h}_{AB}$ is a triplet (i.e., $S_{tot}=1$). The three eigenstates forming this triplet come from the $\nu=0$ sectors of the $m=0,\pm1$ subspaces. For the basis states corresponding to $m=1$, see Table~\ref{tab:m1}. Next in the spectrum we find two more triplets. Of these, one set of triplet comes from $\nu=1$ and $m=0,1,\bar{1}$; the second triplet is formed in the subspaces given by $\nu=\bar{1}$ and $m=0,1,\bar{1}$. Let the 9 eigenstates in these 3 triplets be denoted as $\ket{t_{m\nu}}$. The energy corresponding to $\ket{t_{m0}}$ is denoted as $E_{t0}$ and is shown by red line in~\ref{fig:eig-ABhex}. This energy level remains the second lowest all along $J_B/J_A=0\rightarrow1$. The triplets $\ket{t_{m1}}$ and $\ket{t_{m\bar{1}}}$ are degenerate, and have the energy $E_{t1}$ shown by blue line in Fig.~\ref{fig:eig-ABhex}. The $E_{t1}$ is the third lowest upto $J_B/J_A=0.685$, beyond which another unique singlet becomes lower. This singlet excited state is formed in the $(m,\nu)=(0,0)$ subspace, shown by green line in Fig.~\ref{fig:eig-ABhex}. 

\begin{table}[t]
	\caption{}
	\label{tab:m1}
	\centering 
	\begin{tabular}{c| c} 
		\hline
		 $\nu$ & Basis states for $m=1$ \\ [1ex]
		\hline
		 &~$\frac{1}{\sqrt{3}}(\ket{\uparrow\uparrow\uparrow\downarrow\uparrow\downarrow}+\ket{\uparrow\downarrow\uparrow\downarrow\uparrow\uparrow}+\ket{\uparrow\downarrow\uparrow\uparrow\uparrow\downarrow}$),\\[0.8ex]
		0 &~$\frac{1}{\sqrt{3}}(\ket{\uparrow\uparrow\uparrow\downarrow\downarrow\uparrow}+\ket{\uparrow\downarrow\downarrow\uparrow\uparrow\uparrow}+\ket{\downarrow\uparrow\uparrow\uparrow\uparrow\downarrow}$),\\[0.8ex]
		&~$\frac{1}{\sqrt{3}}(\ket{\uparrow\uparrow\downarrow\uparrow\uparrow\downarrow}+\ket{\downarrow\uparrow\uparrow\downarrow\uparrow\uparrow}+\ket{\uparrow\downarrow\uparrow\uparrow\downarrow\uparrow}$),\\[0.8ex]
		&~$\frac{1}{\sqrt{3}}(\ket{\uparrow\uparrow\downarrow\uparrow\downarrow\uparrow}+\ket{\downarrow\uparrow\downarrow\uparrow\uparrow\uparrow}+\ket{\downarrow\uparrow\uparrow\uparrow\downarrow\uparrow}$),\\[0.8ex]
		&~$\frac{1}{\sqrt{3}}(\ket{\uparrow\uparrow\uparrow\uparrow\downarrow\downarrow}+\ket{\uparrow\uparrow\downarrow\downarrow\uparrow\uparrow}+\ket{\downarrow\downarrow\uparrow\uparrow\uparrow\uparrow})$\\[1.8ex] \hline
		
		&~$\frac{1}{\sqrt{3}}(\omega\ket{\uparrow\uparrow\uparrow\downarrow\uparrow\downarrow}+\omega^2\ket{\uparrow\downarrow\uparrow\downarrow\uparrow\uparrow}+\ket{\uparrow\downarrow\uparrow\uparrow\uparrow\downarrow}$),\\[0.8ex]
		1 &~$\frac{1}{\sqrt{3}}(\omega\ket{\uparrow\uparrow\uparrow\downarrow\downarrow\uparrow}+\omega^2\ket{\uparrow\downarrow\downarrow\uparrow\uparrow\uparrow}+\ket{\downarrow\uparrow\uparrow\uparrow\uparrow\downarrow}$),\\[0.8ex]
		&~$\frac{1}{\sqrt{3}}(\omega\ket{\uparrow\uparrow\downarrow\uparrow\uparrow\downarrow}+\omega^2\ket{\downarrow\uparrow\uparrow\downarrow\uparrow\uparrow}+\ket{\uparrow\downarrow\uparrow\uparrow\downarrow\uparrow}$),\\[0.8ex]
		&~$\frac{1}{\sqrt{3}}(\omega\ket{\uparrow\uparrow\downarrow\uparrow\downarrow\uparrow}+\omega^2\ket{\downarrow\uparrow\downarrow\uparrow\uparrow\uparrow}+\ket{\downarrow\uparrow\uparrow\uparrow\downarrow\uparrow}$),\\[0.8ex]
		&~$\frac{1}{\sqrt{3}}(\omega\ket{\uparrow\uparrow\uparrow\uparrow\downarrow\downarrow}+\omega^2\ket{\uparrow\uparrow\downarrow\downarrow\uparrow\uparrow}+\ket{\downarrow\downarrow\uparrow\uparrow\uparrow\uparrow})$\\[1.8ex] \hline
		
		&~$\frac{1}{\sqrt{3}}(\omega^2\ket{\uparrow\uparrow\uparrow\downarrow\uparrow\downarrow}+\omega\ket{\uparrow\downarrow\uparrow\downarrow\uparrow\uparrow}+\ket{\uparrow\downarrow\uparrow\uparrow\uparrow\downarrow}$),\\[0.8ex]
		$\bar{1}$ &~$\frac{1}{\sqrt{3}}(\omega^2\ket{\uparrow\uparrow\uparrow\downarrow\downarrow\uparrow}+\omega\ket{\uparrow\downarrow\downarrow\uparrow\uparrow\uparrow}+\ket{\downarrow\uparrow\uparrow\uparrow\uparrow\downarrow}$),\\[0.8ex]
		&~$\frac{1}{\sqrt{3}}(\omega^2\ket{\uparrow\uparrow\downarrow\uparrow\uparrow\downarrow}+\omega\ket{\downarrow\uparrow\uparrow\downarrow\uparrow\uparrow}+\ket{\uparrow\downarrow\uparrow\uparrow\downarrow\uparrow}$),\\[0.8ex]
		&~$\frac{1}{\sqrt{3}}(\omega^2\ket{\uparrow\uparrow\downarrow\uparrow\downarrow\uparrow}+\omega\ket{\downarrow\uparrow\downarrow\uparrow\uparrow\uparrow}+\ket{\downarrow\uparrow\uparrow\uparrow\downarrow\uparrow}$),\\[0.8ex]
		&~$\frac{1}{\sqrt{3}}(\omega^2\ket{\uparrow\uparrow\uparrow\uparrow\downarrow\downarrow}+\omega\ket{\uparrow\uparrow\downarrow\downarrow\uparrow\uparrow}+\ket{\downarrow\downarrow\uparrow\uparrow\uparrow\uparrow})$\\[1ex]
		\hline 
		\end{tabular}
		\end{table}
		
Next we derive a representation of the six spins of the hexagon in terms of the singlet ground state and the 3 triplets, i.e. a total of 10 eigenstates: $\{\ket{s}$,$\ket{t_{00}}$,$\ket{t_{10}}$,$\ket{t_{\bar{1}0}}$,$\ket{t_{01}}$,$\ket{t_{11}}$,$\ket{t_{\bar{1}1}}$,$\ket{t_{0\bar{1}}}$,$\ket{t_{1\bar{1}}}$,$\ket{t_{\bar{1}\bar{1}}}\}$. We ignore the singlet excited state mentioned above, because it doesn't form a matrix element with the singlet ground state. We also ignore all the other higher energy eigenstates, because we want to develop a description that is essentially minimal.  

For the ten low-energy eigenstates identified above, we introduce ten bosonic operators as follows.
\begin{align}
	\begin{split}
		\ket{s}&=\hat{s}^\dag \ket{0}\\
		\ket{t_{m\nu}}&=\hat{t}^\dag_{m\nu}\ket{0}
	\end{split}
	\label{eq:toperators}
\end{align}
Here, the creation of a boson by applying $\shat^\dag$ on the vacuum $|0\rangle$ corresponds to having the singlet ground state $|s\rangle$ on the hexagon; likewise for $\that^\dag_{m\nu}$. Since the auxiliary bosonic Fock space is infinite dimensional, the bosons are required to satisfy the constraint, $\shat^\dag\shat + \sum_{m,\nu}\that^\dag_{m\nu}\that^{ }_{m\nu} = 1$, to conform to the dimension of the spin Hilbert space. 

We can write the six spins of a hexagon in terms of these 10 eigenstates. This is a reasonable approximation to formulate an effective low-energy theory. We evaluate the matrix elements of every component of the six spins ($l=1,6$), and write the spin operators in the bra-ket notation. Every term in the bra-ket notation is then made to correspond to a bilinear (of one creation and one annihilation operators) in the bosonic representation. For instance,  $\langle s|S^z_l |t_{m\nu}\rangle |s\rangle\langle t_{m\nu}|$ corresponds to $\langle s|S^z_l |t_{m\nu}\rangle \shat^\dag \that_{m\nu}$. In a physically motivated simplification of this representation, we treat $\shat$ and $\shat^\dag$ in mean-field approximation by the mean singlet amplitude $\sbar$. This $\sbar$ is meant to describe the mean-field hexagonal singlet (HS) state on the full lattice. Finally we keep only those terms which are directly coupled to $\sbar$, i.e. the terms which make the HS state quantum fluctuate directly through triplet excitations. With these simplifications, we get the following triplon representation of the spins on a hexagon. 
\begin{align}
		S_l^z &\approx \bar{s}\left[\mathcal{C}^l_{00}(\that_{00}+\that_{00}^\dag)+\left(\mathcal{C}^l_{01}\that_{01}+\mathcal{C}^{l*}_{01}\that_{0\bar{1}}+{\rm h.c.}\right)\right]\\
		S_l^+ &\approx \bar{s}\left[\mathcal{C}^l_{\bar{1}0}(\that_{\bar{1}0}-\that_{10}^\dag)+\mathcal{C}^l_{\bar{1}1}(\that_{\bar{1}1}-\that^\dag_{1\bar{1}}) +\mathcal{C}^{l*}_{\bar{1}1}(\that_{\bar{1}\bar{1}} -\that_{11}^\dag)\right]
	\label{eq:finalre}
\end{align}
where $\mathcal{C}^l_{00}=\bra{s}S_l^z\ket{t_{00}}$, $\mathcal{C}^l_{01}=\bra{s}S_l^z\ket{t_{01}}$, 
$\mathcal{C}^l_{\bar{1}0}=\bra{s}S_l^+\ket{t_{\bar{1}0}}$, and $\mathcal{C}^l_{\bar{1}1} =\bra{s}S_l^+\ket{t_{\bar{1}1}}$ are the matrix elements in terms of which the other matrix elements can be expressed as $\mathcal{C}^l_{m\nu} = \mathcal{C}^{l*}_{m\bar{\nu}}$ and $\mathcal{C}^{l}_{m\nu} = \mathcal{C}^{l}_{\bar{m}\nu}$. Moreover, the coefficients corresponding to the third and fifth spins are related to that of the first spin as: $\mathcal{C}^3_{m\nu} = \omega^{2\nu} \mathcal{C}^1_{m\nu}$ and $\mathcal{C}^5_{m\nu} = \omega^{\nu} \mathcal{C}^1_{m\nu}$. Similarly, the coefficients corresponding to the fourth and sixth spins are related to that of the second spin as:  
$\mathcal{C}^4_{m\nu} = \omega^{2\nu} \mathcal{C}^2_{m\nu}$ and $\mathcal{C}^6_{m\nu} = \omega^{\nu} \mathcal{C}^2_{m\nu}$.

\begin{table}[t]
	\caption{}
	\label{tab:m2}
	\centering 
	\begin{tabular}{c| c} 
		\hline \\[-1.8ex]
		 $\nu$ & Basis states for $m=2$ \\[0.8ex] \hline \\[-1.8ex]
		0 &~$\frac{1}{\sqrt{3}}(\ket{\uparrow\uparrow\uparrow\uparrow\uparrow\downarrow}+\ket{\uparrow\uparrow\uparrow\downarrow\uparrow\uparrow}+\ket{\uparrow\downarrow\uparrow\uparrow\uparrow\uparrow}$),\\[0.8ex] 
		&~$\frac{1}{\sqrt{3}}(\ket{\uparrow\uparrow\uparrow\uparrow\downarrow\uparrow}+\ket{\uparrow\uparrow\downarrow\uparrow\uparrow\uparrow}+\ket{\downarrow\uparrow\uparrow\uparrow\uparrow\uparrow})$\\[1.2ex] \hline \\[-1.8ex]
		1 &~$\frac{1}{\sqrt{3}}(\omega\ket{\uparrow\uparrow\uparrow\uparrow\uparrow\downarrow}+\omega^2\ket{\uparrow\uparrow\uparrow\downarrow\uparrow\uparrow}+\ket{\uparrow\downarrow\uparrow\uparrow\uparrow\uparrow}$),\\[0.8ex]
		&~$\frac{1}{\sqrt{3}}(\omega\ket{\uparrow\uparrow\uparrow\uparrow\downarrow\uparrow}+\omega^2\ket{\uparrow\uparrow\downarrow\uparrow\uparrow\uparrow}+\ket{\downarrow\uparrow\uparrow\uparrow\uparrow\uparrow})$\\[1.2ex] \hline \\[-1.8ex]
        $\bar{1}$ &~$\frac{1}{\sqrt{3}}(\omega^2\ket{\uparrow\uparrow\uparrow\uparrow\uparrow\downarrow}+\omega\ket{\uparrow\uparrow\uparrow\downarrow\uparrow\uparrow}+\ket{\uparrow\downarrow\uparrow\uparrow\uparrow\uparrow}$),\\[0.8ex]
		&~$\frac{1}{\sqrt{3}}(\omega^2\ket{\uparrow\uparrow\uparrow\uparrow\downarrow\uparrow}+\omega\ket{\uparrow\uparrow\downarrow\uparrow\uparrow\uparrow}+\ket{\downarrow\uparrow\uparrow\uparrow\uparrow\uparrow})$ \\[1.2ex] \hline
		\end{tabular}
		\end{table}
		
Next we describe the quintet ($S_{tot} = 2$) and the heptet ($S_{tot}=3$) eigenstates; they would be required for the theory of magnetization in Sec.~\ref{sec:Mag}. The heptet eigenstates, denoted as $\ket{h_{m0}}$,  are unique and symmetric under rotation. 
The fully polarized $\ket{\uparrow\uparrow\uparrow\uparrow\uparrow\uparrow}$ is the $\ket{h_{30}}$ with eigenvalue $3(J_A+J_B)/4$; the other $\ket{h_{m0}}$ states can be generated from it by the repeated application of $S^-_{tot}$. There are a total of five different quintets denoted as $\ket{q_{m0}}$ with eigenvalue $E_{q0}=-(J_A+J_B)/4$, $\ket{q_{m1}}$ and $\ket{q_{m\bar{1}}}$ with same eigenvalue $E_{q1}=[3 ( J_A + J_B ) - \sqrt{17 J_A^2 - 14 J_A J_B + 17 J_B^2} ]/8 $, and $\ket{q^\prime_{m1}}$ and $\ket{q^\prime_{m\bar{1}}}$ with eigenvalues $E^\prime_{q1}=[3 ( J_A + J_B ) + \sqrt{17 J_A^2 - 14 J_A J_B + 17 J_B^2} ]/8 $. Of these, the $m=2$ states can be written in terms of the basis states given in Table~\ref{tab:m2}. For instance, $\ket{q_{20}}=(\ket{\uparrow\uparrow\uparrow\uparrow\uparrow\downarrow}+\ket{\uparrow\uparrow\uparrow\downarrow\uparrow\uparrow}+\ket{\uparrow\downarrow\uparrow\uparrow\uparrow\uparrow}-\ket{\uparrow\uparrow\uparrow\uparrow\downarrow\uparrow}-\ket{\uparrow\uparrow\downarrow\uparrow\uparrow\uparrow}-\ket{\downarrow\uparrow\uparrow\uparrow\uparrow\uparrow})/\sqrt{6}$ is an anti-symmetric linear superposition of the two states in the $\nu=0$ sector; the $m=2$ eigenstates for $\nu=1,\bar{1}$ can be obtained from the corresponding $2\times2$ matrices for $\hat{h}_{AB}$.		

\section{Hamiltonian matrix and Bogoliubov diagonalization for the HS state triplon dynamics}{\label{app:HShamil}}
The $\mathcal{H}_{\vec{k}}$ in Eq.~\eqref{eq:tHS} is an $18\cross18$ matrix in the Nambu basis. We can write it as,
$	\mathcal{H}_{\vec{k}}=\begin{pmatrix}
		\mathcal{M}_{\vec{k}} & & \mathcal{W}_{\vec{k}}\\ & & \\
		\mathcal{W}_{\vec{k}}^\dag & & \mathcal{M}_{-\vec{k}}^*
	\end{pmatrix}$,
where $\mathcal{M}_{\vec{k}}$ and $\mathcal{W}_{\vec{k}}$ are two $9\cross9$ matrices given below.
\begin{widetext}
	\begin{subequations}
		\begin{equation}
			\mathcal{M}_{\vec{k}}  =  \begin{pmatrix}
				&D_{00}^{\vec{k}}&0&0&A_{0001}^{\vec{k}}&0&0&A_{0001}^{-\vec{k}*}&0&0\\
				&0&D_{00}^{\vec{k}}&0&0&A_{1011}^{\vec{k}}&0&0&A_{1011}^{-\vec{k} *}&0\\
				&0&0&D_{00}^{\vec{k}}&0&0&A_{1011}^{\vec{k}}&0&0&A_{1011}^{-\vec{k}*}\\
				&A_{0001}^{\vec{k}*}&0&0&D_{01}^{\vec{k}}&0&0&A_{010\bar{1}}^{\vec{k}}&0&0\\
				&0&A_{1011}^{\vec{k} *}&0&0&D_{11}^{\vec{k}}&0&0&A_{111\bar{1}}^{\vec{k}}&0\\
				&0&0&A_{1011}^{\vec{k}*}&0&0&D_{11}^{\vec{k}}&0&0&A_{111\bar{1}}^{\vec{k}}\\
				&A_{0001}^{-\vec{k}}&0&0&A_{010\bar{1}}^{\vec{k}*}&0&0&D_{01}^{\vec{k}}&0&0\\
				&0&A_{1011}^{-\vec{k}}&0&0&A_{111\bar{1}}^{\vec{k}*}&0&0&D_{11}^{\vec{k}}&0\\
				&0&0&A_{1011}^{-\vec{k}}&0&0&A_{111\bar{1}}^{\vec{k}*}&0&0&D_{11}^{\vec{k}}	
			\end{pmatrix} 
		\end{equation}
		\begin{equation}
			\mathcal{W}_{\vec{k}}  =  \begin{pmatrix}
				&B_{00}^{\vec{k}}&0&0&A_{0001}^{-\vec{k}*}&0&0&A_{0001}^{\vec{k}}&0&0\\
				&0&0&B_{10\bar{1}0}^{\vec{k}}&0&0&-A_{1011}^{-\vec{k}*}&0&0&-A_{1011}^{\vec{k}}\\
				&0&B_{10\bar{1}0}^{-\vec{k}}&0&0&-A_{1011}^{-\vec{k}*}&0&0&-A_{1011}^{\vec{k}}&0\\
				&A_{0001}^{\vec{k}*}&0&0&B_{0\bar{1}}^{\vec{k}*}&0&0&B_{010\bar{1}}^{\vec{k}}&0&0\\
				&0&0&-A_{1011}^{\vec{k}*}&0&0&-A_{111\bar{1}}^{\vec{k}}&0&0&B_{11\bar{1}\bar{1}}^{\vec{k}}\\
				&0&-A_{1011}^{\vec{k}*}&0&0&-A_{111\bar{1}}^{-\vec{k}}&0&0&B_{11\bar{1}\bar{1}}^{\vec{k}}&0\\
				&A_{0001}^{-\vec{k}}&0&0&B_{010\bar{1}}^{-\vec{k}}&0&0&B_{0\bar{1}}&0&0\\
				&0&0&-A_{1011}^{-\vec{k}}&0&0&B_{11\bar{1}\bar{1}}^{-\vec{k}}&0&0&-A_{111\bar{1}}^{\vec{k}*}\\
				&0&-A_{1011}^{-\vec{k}}&0&0&B_{11\bar{1}\bar{1}}^{-\vec{k}}&0&0&-A_{111\bar{1}}^{-\vec{k}*}&0
			\end{pmatrix}
		\end{equation}
	\end{subequations}
\end{widetext}
The elements of these matrices are given as follows:
\begin{subequations}
\begin{align}
	\begin{split}
		D_{00}^{\vec{k}}&= \frac{\lambda+ E_{t0}}{2} + J_C \bar{s}^2\mathcal{C}_{00}^1\mathcal{C}_{00}^2f_{\vec{k}}^0\\
		D_{01}^{\vec{k}}&= \frac{\lambda+ E_{t1}}{2} + J_C\bar{s}^2 Re(\omega\mathcal{C}_{01}^1\mathcal{C}_{01}^{2*}\gamma_{\vec{k}}^0) \\
		D_{11}^{\vec{k}}&= \frac{\lambda +E_{t1}}{2} + \frac{J_C\bar{s}^2}{2} Re(\omega\mathcal{C}_{\bar{1}1}^1\mathcal{C}_{\bar{1}1}^{2*}\gamma_{\vec{k}}^0)
		\end{split} 
		\end{align}
		\begin{align}
		\begin{split}
		A_{0001}^{\vec{k}}&= \frac{J_C\bar{s}^2}{2}(\mathcal{C}_{00}^1\mathcal{C}_{01}^2\gamma_{\vec{k}} +\omega\mathcal{C}_{01}^1\mathcal{C}_{00}^2\gamma_{-\vec{k}} )\\
		A_{1011}^{\vec{k}}&= \frac{J_C\bar{s}^2}{4}(\mathcal{C}_{\bar{1}0}^1\mathcal{C}_{\bar{1}1}^2\gamma_{\vec{k}} +\omega\mathcal{C}_{\bar{1}1}^1\mathcal{C}_{\bar{1}0}^2\gamma_{-\vec{k}} )\\
		A_{010\bar{1}}^{\vec{k}}&=J_C\bar{s}^2 (\mathcal{C}_{01}^{1}\mathcal{C}_{01}^{2}f_{\vec{k}})^*\\
		A_{111\bar{1}}^{\vec{k}}&=\frac{J_C\bar{s}^2}{2} (\mathcal{C}_{\bar{1}1}^{1}\mathcal{C}_{\bar{1}1}^{2}f_{\vec{k}})^*
		\end{split}
		\end{align}
		\begin{align}
		\begin{split}
		B_{00}^{\vec{k}}&= J_C\bar{s}^2\mathcal{C}_{00}^{1}\mathcal{C}_{00}^{2}f_{\vec{k}}^0\\
		B_{0\bar{1}}^{\vec{k}}&=J_C \bar{s}^2\mathcal{C}_{01}^1\mathcal{C}_{01}^2 f_{\vec{k}}\\
		B_{10\bar{1}0}^{\vec{k}}&= - \frac{J_C\bar{s}^2}{2}\mathcal{C}_{\bar{1}0}^{1}\mathcal{C}_{\bar{1}0}^{2}f_{\vec{k}}\\
		B_{010\bar{1}}^{\vec{k}}&=J_C\bar{s}^2 Re(\omega\mathcal{C}_{01}^1\mathcal{C}_{01}^{2*}\gamma_{\vec{k}}^0)\\
		B_{11\bar{1}\bar{1}}^{\vec{k}}&=-\frac{J_C\bar{s}^2}{2} Re(\omega\mathcal{C}_{\bar{1}1}^1\mathcal{C}_{\bar{1}1}^{2*}\gamma_{\vec{k}}^0)
	 \end{split}
\end{align}
\end{subequations}
where
\begin{align}
	\begin{split}
		f_{\vec{k}}^0&=\cos{k_1}+\cos{k_2}+\cos{k_3}\\
		f_{\vec{k}}&=\cos{k_1}+\omega^2 \cos{k_2}+\omega \cos{k_3}\\
		\gamma_{\vec{k}}^0&=e^{-i k_1}+e^{-i k_2}+e^{i k_3}\\
		\gamma_{\vec{k}}&=\omega e^{i k_1}+\omega^2e^{i k_2}+e^{-i k_3}
	\end{split}
\end{align}
for $k_1$, $k_2$, $k_3$ defined in the main text [see below Eq.~\eqref{eq:nuk}]. 

To diagonalize the triplon Hamiltonian $H_{tHS}$ of Eq.~\eqref{eq:tHS}, as per the prescription due to Bogoliubov, we first multiply $\mathcal{H}_{\vec{k}}$ with the matrix
\begin{equation}
	\centering
	\Lambda=	\begin{pmatrix}
		&\mathbb{I}_{9} & 0\\
		&0 & -\mathbb{I}_{9}
	\end{pmatrix}
	\label{eq:matlambda}
\end{equation}
from the left hand side; here $\mathbb{I}_{9}$ is a $9\cross9$ identity matrix. We then diagonalise the matrix $\Lambda\mathcal{H}_{\vec{k}}$. Its eigenvalues come in pairs, i.e. for every positive eigenvalue there occurs a negative eigenvalue with same magnitude. Of these, the positive eigenvalues are the triplon dispersions $\epsilon_{i\vec{k}}$ in Eq.~\eqref{eq:HSEg}. 

\section{Spin-wave analysis of the ABC model}\label{app:SW}
Consider the perfect N\'eel antiferromagnetic state on the kagome-honeycomb lattice. In a unit-cell (say, AB-hexagon) at position $\vec{R}$, the odd-numbered spins, assumed to be aligned in the $+z$ direction, can be written in the Holstein-Primakoff representation as
\begin{subequations}
\begin{align}
S_{1,\vec{R}}^z = S - \ahat^\dag_{1,\vec{R}} \ahat^{ }_{1,\vec{R}}\,,~ S_{1,\vec{R}}^+ \approx \sqrt{2S} \ahat^{ }_{1,\vec{R}}
\label{eq:HPodd}
\end{align}
and likewise for $\vec{S}_{3,\vec{R}}$ and $\vec{S}_{5,\vec{R}}$. Correspondingly, the even-numbered spins are pointed along $-z$ direction. Hence, in the Holstein-Primakoff representation, 
\begin{align}
S_{2,\vec{R}}^z = -S + \ahat^\dag_{2,\vec{R}} \ahat^{ }_{2,\vec{R}}\,,~S_{2,\vec{R}}^+ \approx \sqrt{2S} \ahat_{2,\vec{R}}^\dag
	\label{eq:HPeven}
\end{align} 
\end{subequations}
and likewise for $\vec{S}_{4,\vec{R}}$ and $\vec{S}_{6,\vec{R}}$. We apply this to the ABC model [Eq.~\eqref{eq:HABC}], together with the Fourier transformation, $\ahat_{l,\vec{R}} = \frac{1}{\sqrt{N_{uc}}}\sum_{\vec{k}} e^{i\vec{k}\cdot\vec{R}}\ahat_{l,\vec{k}}$ for $l=1$ to $6$. We finally get the following spin-wave Hamiltonian: 
\begin{equation}
	\Hhat^{ }_{SW} = -3S(S+1)(J_A+J_B+J_C)N_{uc} + \frac{S}{2}\sum_{\vec{k}}\Phi_{\vec{k}}^\dag h_{\vec{k}}\Phi_{\vec{k}}
\end{equation}
where $\Phi^\dag_{\vec{k}}= \begin{pmatrix} \ahat^\dag_{1,\vec{k}} & \ahat^\dag_{2,\vec{k}} & \cdots &  \ahat^\dag_{6,\vec{k}} & \ahat^{ }_{1,-\vec{k}} & \ahat^{ }_{2,-\vec{k}} & \cdots & \ahat^{ }_{6,-\vec{k}}\end{pmatrix} $ is a Nambu row vector, and 
$	h_{\vec{k}}=	\begin{pmatrix}
		A_{\vec{k}} & B_{\vec{k}}\\
		B_{\vec{k}} & A_{\vec{k}}
	\end{pmatrix}$
is a $12\cross12$ matrix with $\bold{A}_{\vec{k}} = (J_A+J_B+J_C) \, \mathbb{I}_{6}$ and 
\begin{align}
	\bold{B}_{\vec{k}} =	\begin{pmatrix}
	0 & J_B &0&J_C e^{i k_2}&0&J_A\\
	J_B & 0 &J_A&0 &J_C e^{i k_3}&0\\
	0 & J_A &0&J_B &0&J_Ce^{i k_1}\\
	J_Ce^{-i k_2} &0& J_B &0&J_A&0\\
	0 & J_Ce^{-i k_3} &0&J_A &0&J_B\\
	J_A & 0 &J_C e^{-i k_1}&0&J_B &0
    \end{pmatrix}
	\label{eq:ABk}
\end{align}
for the same $k_1,k_2$ and $k_3$ as defined near Eq.~\eqref{eq:nuk}. By doing Bogoliubov diagonlization of $\Hhat_{SW}$, we get six spin-wave dispersions, $E_{l\vec{k}}$, and the following expression for the ground state energy per unit-cell.
\begin{align}
		E_{gSW} = -3S(S+1)(J_A+J_B+J_C) + \frac{S}{2N_{uc}}\sum_{\vec{k}}\sum_{l=1}^{6}E_{l\vec{k}}
\end{align}
Using this, we calculate the spin-wave energy of the ABC model for $S=1/2$. 

\section{Hamiltonian matrix and other details concerning the theory of 1/3 plateau}\label{app:1by3}
The spins of an AB-hexagon can be represented in the reduced subspace, $\{\ket{s},\ket{t_{10}},\ket{q_{20}},\ket{q_{21}},\ket{q_{2\bar{1}}}\}$, as:
\begin{subequations}
\label{eq:rep1by3}
\begin{align}
S_{l,\vec{R}}^z \approx & ~ C_{1010}^l \, \bar{t}^2 + C_{2020}^l \, \hat{q}^\dag_{20,\vec{R}}\hat{q}_{20,\vec{R}} \nonumber \\
& ~  + C_{2121}^l \Big(\hat{q}^\dag_{21,\vec{R}}\hat{q}_{21,\vec{R}} + \hat{q}^\dag_{2\bar{1},\vec{R}}\hat{q}_{2\bar{1},\vec{R}}\Big)\\
S_{l,\vec{R}}^+ \approx & ~ \bar{t} \Big(C_s^l \hat{s}_{\vec{R}} + C_{20}^l \hat{q}_{20,\vec{R}}^\dag + C_{21}^l \hat{q}_{21,\vec{R}}^\dag + C_{21}^{l*}\hat{q}_{2\bar{1},\vec{R}}^\dag\Big)	
\end{align}
\end{subequations}
where $l=1$ to 6 is the spin label, and the coefficients $C_{20}^l$, $C_{21}^l$, $C_{2020}^l$ etc are the matrix elements defined below.
\begin{align*}
	\begin{split}
		C_{1010}^l=\bra{t_{10}}S_l^z\ket{t_{10}}, &~~ C_{2020}^l=\bra{q_{20}}S_l^z\ket{q_{20}}\\ C_{2121}^l=\bra{q_{21}}S_l^z\ket{q_{21}}, &~~ C_{s}^l=\bra{t_{10}}S_l^+\ket{s}\\
		C_{20}^l=\bra{q_{20}}S_l^+\ket{t_{10}},   &~~ C_{21}^l=\bra{q_{21}}S_l^+\ket{t_{10}}
	\end{split}
\end{align*}
Moreover, $C_{1010}^1=C_{1010}^3=C_{1010}^5$ and $C_{1010}^2=C_{1010}^4=C_{1010}^6$; same is true for $C_{2020}^l$ and $C_{s}^l$. These are real coefficients. The complex coefficients are: $C_{21}^3=\omega C_{21}^1$, $C_{21}^5=\omega^2 C_{21}^1$  and $C_{21}^4=\omega C_{21}^2$, $C_{21}^6=\omega^2 C_{21}^2$. 
 
The constant term, $\epsilon_0^{(\frac{1}{3})}$, in  Eq.~\ref{eq:H1by3} is given by
\begin{subequations}
\begin{align}
		\epsilon_0^{(\frac{1}{3})} =&~ (E_{t0}-h_{ext} + \lambda)\bar{t}^2 - \frac{3}{2}J_C\bar{t}^2 (\chi^{(\frac{1}{3})} - 2\bar{t}^2 C_{1010}^1C_{1010}^2) \nonumber \\
		& -\frac{1}{2}(E_s+E_{q0}+2E_{q1}) + 3(h_{ext}-\lambda) \\
\chi^{(\frac{1}{3})} =&~ C_{1010}^1(C_{2020}^2 + 2C_{2121}^2) + C_{1010}^2(C_{2020}^1 + 2C_{2121}^1) \\
\chi &=  \chi^{(\frac{1}{3})}  - 4 \bar{t}^2 C_{1010}^1C_{1010}^2 \label{eq:chi1by3}
\end{align}
\end{subequations}
and the $8 \times 8$ Hamiltonian matrix in the Nambu basis can be written as
 $\mathcal{H}_{\vec{k}}^{(\frac{1}{3})}=\begin{pmatrix}
	\mathcal{M}_{\vec{k}}^{(\frac{1}{3})} & \mathcal{W}_{\vec{k}}^{(\frac{1}{3})} \\ & \\
	\mathcal{W}_{\vec{k}}^{(\frac{1}{3})\dag} & \mathcal{M}_{-\vec{k}}^{(\frac{1}{3})*}
\end{pmatrix}$  
with

\begin{subequations}
\begin{align}
	\mathcal{M}_{\vec{k}}^{(\frac{1}{3})}  & =  \begin{pmatrix}
		&D_{s}^{\vec{k}}&0&0&0\\
		&0&D_{20}^{\vec{k}}&A_{2021}^{\vec{k}}&A_{2021}^{-\vec{k}*}\\
		&0&A_{2021}^{\vec{k}*}&D_{21}^{\vec{k}}&A_{212\bar{1}}^{\vec{k}}\\
		&0&A_{2021}^{-\vec{k}}&A_{212\bar{1}}^{\vec{k}*}&D_{2\bar{1}}^{\vec{k}}
	\end{pmatrix} \\
	& \nonumber \\
	\mathcal{W}_{\vec{k}}^{(\frac{1}{3})}  & =  \begin{pmatrix}
		&0&B_{s,20}^{\vec{k}}&B_{s,2\bar{1}}^{-\vec{k}*}&B_{s,2\bar{1}}^{\vec{k}}\\
		&B_{s,20}^{-\vec{k}}&0&0&0\\
		&B_{s,2\bar{1}}^{\vec{k}*}&0&0&0\\
		&B_{s,2\bar{1}}^{-\vec{k}}&0&0&0
	\end{pmatrix} 
\end{align}
\end{subequations}
where 
		\begin{eqnarray*}
			D_s^{\vec{k}} &=& \frac{\lambda+E_s}{2} + \frac{J_C\bar{t}^2}{2}C_s^1C_s^2f_{\vec{k}}^0\\
			D_{20}^{\vec{k}} &=& \frac{\lambda+E_{q0}-2h_{ext}}{2} + \frac{J_C\bar{t}^2}{2}C_{20}^1C_{20}^2f_{\vec{k}}^0 \\
			&& +\frac{3J_C\bar{t}^2}{2}(C_{1010}^1C_{2020}^2+C_{2020}^1C_{1010}^2)\\
			D_{21}^{\vec{k}} &=& \frac{\lambda+E_{q1}-2h_{ext}}{2} + \frac{J_C\bar{t}^2}{2}Re(\omega^2C_{21}^1C_{21}^{2*}\gamma_{-\vec{k}}^0)\\
			&& +\frac{3J_C\bar{t}^2}{2}(C_{1010}^1C_{2121}^2 + C_{2121}^1C_{1010}^2)\\
			D_{2\bar{1}}^{\vec{k}} &=& \frac{\lambda+E_{q1}-2h_{ext}}{2} + \frac{J_C\bar{t}^2}{2}Re(\omega C_{21}^{1*}C_{21}^{2}\gamma_{-\vec{k}}^0)\\
			&& +\frac{3J_C\bar{t}^2}{2}(C_{1010}^1C_{2121}^2 + C_{2121}^1C_{1010}^2)\\
			A_{2021}^{\vec{k}} &=& \frac{J_C\bar{t}^2}{4}(C_{20}^1C_{21}^{2*}\gamma_{\vec{k}} + C_{21}^{1*}C_{20}^2\gamma_{-\vec{k}}\omega)\\
			A_{212\bar{1}}^{\vec{k}} &=& \frac{J_C\bar{t}^2}{2}C_{21}^1C_{21}^2f_{\vec{k}}\\
			B_{s,20}^{\vec{k}} &=& \frac{J_C\bar{t}^2}{4}(C_{s}^1C_{20}^2\gamma_{-\vec{k}}^0 + C_{20}^1C_{s}^2\gamma_{\vec{k}}^0)\\
			B_{s,2\bar{1}}^{\vec{k}} &=& \frac{J_C\bar{t}^2}{4}(C_{s}^1C_{21}^{2*}\gamma_{\vec{k}} + C_{21}^{1*}C_{s}^2\gamma_{-\vec{k}}\omega)
		\end{eqnarray*}		

\section{Details of the theory of 2/3 plateau}\label{app:2by3}
The simplified representation of the spins in a hexagonal unit-cell in the subspace, $\{\ket{t_{10}},\ket{q_{20}},\ket{h_{30}}\}$, relevant for 2/3 plateau. 
\begin{align}
	\begin{split}
		S_{l,\vec{R}}^z &= C_{2020}^l \bar{q}^2 + C_{1010}^l \hat{t}_{10,\vec{R}}^\dag\hat{t}_{10,\vec{R}} + C_{3030}^l \hat{h}_{30,\vec{R}}^\dag\hat{h}_{30,\vec{R}}\\
		S_{l,\vec{R}}^+ &= \bar{q} (C_{10}^l \hat{t}_{10,\vec{R}} + C_{30}^l \hat{h}_{30,\vec{R}}^\dag)
	\end{split}
	\label{eq:rep2by3mini}	
\end{align}	
Here,  $C_{1010}^l$, $C_{2020}^l$ and $C_{10}^l$ are same as defined in Appendix~\ref{app:1by3}. Moreover, $C_{3030}^l=\bra{h_{30}}S_l^z\ket{h_{30}}=\frac{1}{2}~\mbox{for}~l=1~\mbox{to}~6$, and $C_{30}^l=\bra{h_{30}}S_l^+\ket{q_{20}}$ take the following values:  $C_{30}^1=C_{30}^3=C_{30}^5=\frac{1}{\sqrt{6}}$ and $C_{30}^2=C_{30}^4=C_{30}^6=-\frac{1}{\sqrt{6}}$.

The constant term and the coefficients in $\Hhat^{(\frac{2}{3})}$ are:
\begin{subequations}
\label{eq:DDFeps}
\begin{align}
D_{t,\vec{k}} =& ~ \lambda + E_{t0} - h_{ext} + J_C \bar{q}^2 C_{10}^1C_{10}^2 f_{\vec{k}}^0 \nonumber \\
		& ~ + 3J_C \bar{q}^2 (C_{2020}^1 C_{1010}^2 + C_{1010}^1C_{2020}^2)\\
D_{h,\vec{k}} =& ~ \lambda + E_{h0} - 3h_{ext} + J_C \bar{q}^2 C_{30}^1C_{30}^2f_{\vec{k}}^0 \nonumber \\
		& ~ + 3J_C \bar{q}^2 (C_{2020}^1 C_{3030}^2 + C_{3030}^1C_{2020}^2)\\
F_{\vec{k}} =& ~  \frac{J_C\bar{q}^2}{2} |C_{10}^1C_{30}^2 \gamma_{\vec{k}}^0 + C_{30}^1C_{10}^2\gamma_{-\vec{k}}^0| \\
\epsilon_0^{(\frac{2}{3})} =&~ (E_{q0}-2h_{ext} +\lambda)\bar{q}^2  + 3J_C\bar{q}^4 \, C_{2020}^1C_{2020}^2 \nonumber \\
&  - 3J_C\bar{q}^2(C_{2020}^1C_{3030}^2 + C_{3030}^1 C_{2020}^2) \nonumber \\
& -E_{h0}+ 3h_{ext} -2\lambda \\
\tilde{\chi} =& ~ 2\bar{q}^2C_{2020}^1C_{2020}^2 - (C_{2020}^1C_{3030}^2 + C_{3030}^1C_{2020}^2)
\end{align}
\end{subequations}

\bibliography{references.bib}
\end{document}